\newif
\ifconfver
\confvertrue        
\documentclass[preprint,12pt]{elsarticle}

\usepackage{xspace,empheq,fancybox,amssymb,amsfonts,graphicx,epstopdf,epsfig,syntonly,times,subfigure} 
\usepackage{caption}
\usepackage{psfrag,color,bm,array,tabularx}
\usepackage{url,footnote,xspace,syntonly,algorithmic}
\usepackage{verbatim,multirow}
\usepackage[linesnumbered,ruled,vlined]{algorithm2e}
\usepackage{nomencl}
\usepackage{makecell}
\makenomenclature

\usepackage{graphicx}

\usepackage{textcomp}
\usepackage{placeins}
\usepackage{xcolor}
\usepackage{mathtools}
\usepackage{bbm, booktabs}
\usepackage{amsmath, amsthm, lipsum}       
\usepackage{ifthen,etoolbox}
\usepackage{array}
\usepackage{tabularx}
\usepackage{amssymb}
\usepackage{pifont}
\usepackage[table]{xcolor}
\newcommand{\xmark}{\ding{55}}

\newcolumntype{C}[1]{>{\centering\arraybackslash}p{#1}}
\newcolumntype{R}[1]{>{\raggedleft\arraybackslash}p{#1}}

\newcolumntype{M}[1]{>{\centering\arraybackslash}m{#1}}
\newcolumntype{N}{@{}m{0pt}@{}}

\DeclareMathOperator
*{\argmax}{\arg\max}

\def\BibTeX{{\rm B\kern-.05em{\sc i\kern-.025em b}\kern-.08em
		T\kern-.1667em\lower.7ex\hbox{E}\kern-.125emX}}

\newtheorem{remark}{\bfseries Remark}
\input{mysymbol.sty}
\allowdisplaybreaks

\newcommand{\KB}{\color{black}{}}
\newcommand{\KBA}{\color{black}{}}
\newcommand{\KBB}{\color{black}{}}

\renewcommand{\nomgroup}[1]{%
  \ifthenelse{\equal{#1}{A}}{\item[\textbf{Indices}]\vspace{8pt}}{
  \ifthenelse{\equal{#1}{B}}{\vspace{10pt}\item[\textbf{Sets}]\vspace{5pt}}{
  \ifthenelse{\equal{#1}{C}}{\vspace{10pt}\item[\textbf{Variables}]\vspace{5pt}}{
  \ifthenelse{\equal{#1}{D}}{\vspace{10pt}\item[\textbf{Parameters}]\vspace{5pt}}{}}}}%
}

\begin{document}
	


\begin{frontmatter}

\title{Stochastic Multi-Segment Scheduling of Variable-Speed Pumped Storage Hydropower for Energy and Ancillary Services Provision}

\fntext[fn1]{This work was supported by Chungnam National University.}

\author[inst1]{Kyung-bin Kwon}
\ead{kyung-bin.kwon@pnnl.gov}

\author[inst2]{SangWoo Park}
\ead{sangwoo.park@njit.edu}

\author[inst3]{Dam Kim\corref{cor1}}
\ead{damkim@cnu.ac.kr}

\cortext[cor1]{Corresponding author.}

\address[inst1]{Optimization and Control Group, Pacific Northwest National Laboratory, Richland, WA 99352, USA}
\address[inst2]{Department of Mechanical and Industrial Engineering, New Jersey Institute of Technology, Newark, NJ 07103, USA}
\address[inst3]{Department of Convergence System Engineering, Chungnam National University, Daejeon 34134, South Korea}

\begin{abstract}
Variable-speed pumped storage hydropower (VS-PSH) offers long-duration energy storage alongside ancillary services in competitive electricity markets. However, its operation and scheduling are challenged by head-dependent nonlinearities, discrete mode transitions, and energy-continuity constraints. This study proposes a stochastic framework for VS-PSH that employs a multi-segment bidding structure to generate market-consistent energy and synchronized reserve offers in compliance with market rules. The framework explicitly incorporates physical constraints, including head-dependent capability limits, discrete pumping and generating modes, as well as state-of-charge (SoC) and head dynamics, within a stochastic mixed-integer linear programming (MILP) formulation. Price uncertainty is represented through a scenario-based modeling approach that scales base-case prices and allows variations in (dis)charging incentives. The stochastic MILP produces optimal energy and mode schedules that maintain feasible SoC trajectories across scenarios and ensure physically feasible operating strategies. Case studies under different levels of price variability demonstrate the operational feasibility and market applicability of the proposed framework, showing effective coordination between energy arbitrage and reserve provision under uncertainty. These results highlight the operational and economic value of VS-PSH as a grid-scale energy storage resource.
\end{abstract}

\begin{keyword}
Variable-speed pumped storage hydropower (VS-PSH), multi-segment scheduling, stochastic optimization, co-optimization, energy storage resource (ESR) participation
\end{keyword}

\end{frontmatter}


\printnomenclature

\nomenclature[A]{\(t\)}{Index for time intervals (hour index).}
\nomenclature[A]{\(s\)}{Index for price scenarios.}
\nomenclature[A]{\(i, j\)}{Indices for bid segments.}
\nomenclature[A]{\(k\)}{Index for head/discretization levels.}

\nomenclature[B]{\(\mathcal{T}\)}{Set of all time intervals indexed by \(t\).}
\nomenclature[B]{\(\mathcal{S}\)}{Set of all price scenarios indexed by \(s\).}
\nomenclature[B]{\(\mathcal{I}\)}{Set of bid segments (index set for \(i, j\)).}
\nomenclature[B]{\(\mathcal{K}\)}{Set of head/discretization levels (index set for \(k\)).}
\nomenclature[B]{\(\mathcal{F}^{ch}(h)\)}{Feasible set of charging power as a function of hydraulic head \(h\); represented by linear upper/lower bounding lines. (MW)}
\nomenclature[B]{\(\mathcal{F}^{ds}(h)\)}{Feasible set of discharging power as a function of hydraulic head \(h\); represented by linear upper/lower bounding lines. (MW)}
\nomenclature[B]{\(\mathcal{I}^{ch}_1,\mathcal{I}^{ch}_2,\mathcal{I}^{ds}_1,\mathcal{I}^{ds}_2\)}{Index subsets for the linear inequalities that define upper/lower bounding lines in \(\mathcal{F}^{ch}(h)\) and \(\mathcal{F}^{ds}(h)\).}

\nomenclature[C]{\(x^{\mathrm{ch}}_{t}\)}{Binary — 1 if unit is in charging mode at time \(t\), 0 otherwise.}
\nomenclature[C]{\(x^{\mathrm{ds}}_{t}\)}{Binary — 1 if unit is in discharging mode at time \(t\), 0 otherwise.}
\nomenclature[C]{\(x^{\mathrm{con}}_{t}\)}{Binary — 1 if unit is in continuous (both) mode at time \(t\), 0 otherwise.}
\nomenclature[C]{\(x^{\mathrm{non}}_{t}\)}{Binary — 1 if unit is non-operational at time \(t\), 0 otherwise.}
\nomenclature[C]{\(z^{ch}_{i,t,s}\)}{Binary — 1 if charging bid segment \(i\) is selected/cleared at time \(t\) in scenario \(s\).}
\nomenclature[C]{\(z^{ds}_{i,t,s}\)}{Binary — 1 if discharging bid segment \(i\) is selected/cleared at time \(t\) in scenario \(s\).}
\nomenclature[C]{\(z^{syn}_{i,t,s}\)}{Binary — 1 if synchronized-reserve bid segment \(i\) is selected/cleared at time \(t\) in scenario \(s\).}
\nomenclature[C]{\(\hat{z}^{ch}_{i,t,s}\)}{Auxiliary binary — 1 if charging segment \(i\) is price-competitive at \(t,s\).}
\nomenclature[C]{\(\hat{z}^{ds}_{i,t,s}\)}{Auxiliary binary — 1 if discharging segment \(i\) is price-competitive at \(t,s\).}
\nomenclature[C]{\(\hat{z}^{syn}_{i,t,s}\)}{Auxiliary binary — 1 if synchronized-reserve segment \(i\) is price-competitive at \(t,s\).}
\nomenclature[C]{\(\gamma^{ds}_{t,s}\)}{Binary — sign-regime indicator for energy clearing price at \(t,s\): 1 if \(\delta^{en}_{t,s}\ge 0\) (discharge regime), 0 if \(\delta^{en}_{t,s}<0\) (charge regime).}
\nomenclature[C]{\(\gamma^{syn}_{t,s}\)}{Binary — sign/feasibility indicator for synchronized-reserve clearing at \(t,s\).}
\nomenclature[C]{\(\rho^{syn}_{t,s}\)}{Continuous in \([0,1]\) — utilization ratio of synchronized reserve capacity actually deployed in scenario \(s\) at time \(t\).}
\nomenclature[C]{\(b^{ch}_{t,s}\)}{Accepted charging power (negative by sign convention) at time \(t\) under scenario \(s\). (MW)}
\nomenclature[C]{\(b^{ds}_{t,s}\)}{Accepted discharging power at time \(t\) under scenario \(s\). (MW)}
\nomenclature[C]{\(b^{syn}_{t,s}\)}{Cleared synchronized reserve capacity at time \(t\) under scenario \(s\). (MW)}
\nomenclature[C]{\(b_{t,s}\)}{Net accepted power at time \(t\) under scenario \(s\). (MW)}
\nomenclature[C]{\(C_{t,s}\)}{State of charge (SoC) of the reservoir at time \(t\) under scenario \(s\). }
\nomenclature[C]{\(h_{t,s}\)}{Hydraulic head at time \(t\) under scenario \(s\).}
\nomenclature[C]{\(\delta^{en}_{t,s}\)}{Energy-market clearing price at time \(t\) under scenario \(s\). (\$/MWh)}
\nomenclature[C]{\(\delta^{syn}_{t,s}\)}{Synchronized-reserve clearing price at time \(t\) under scenario \(s\). (\$/MWh)}
\nomenclature[C]{\(\hat{p}^{ch}_{i,t}\)}{Offered price for charging bid segment \(i\) at time \(t\). (\$/MWh)}
\nomenclature[C]{\(\hat{p}^{ds}_{i,t}\)}{Offered price for discharging bid segment \(i\) at time \(t\). (\$/MWh)}
\nomenclature[C]{\(\hat{p}^{syn}_{i,t}\)}{Offered price for synchronized-reserve bid segment \(i\) at time \(t\). (\$/MWh)}

\nomenclature[D]{\(\hat{b}^{ch}_{i},\,\hat{b}^{ds}_{i}\)}{Discretized charging and discharging power levels for bid segment \(i\). (MW)}
\nomenclature[D]{\(\underline{b}^{ch}_{t,s},\,\bar{b}^{ch}_{t,s}\)}{Minimum and maximum allowable charging power at time \(t\) under scenario \(s\) (from \(\mathcal{F}^{ch}(h_{t,s})\)). (MW)}
\nomenclature[D]{\(\underline{b}^{ds}_{t,s},\,\bar{b}^{ds}_{t,s}\)}{Minimum and maximum allowable discharging power at time \(t\) under scenario \(s\) (from \(\mathcal{F}^{ds}(h_{t,s})\)). (MW)}
\nomenclature[D]{\(\eta\)}{Charge/discharge efficiency factor (round-trip modeling: charging loss and discharging loss). (pu)}
\nomenclature[D]{\({\KBB M^{ch},\, M^{ds}}\)}{{\KBB Tightened Big-M constants for charging and discharging bid-acceptance constraints.}}
\nomenclature[D]{\({\KBB M^{syn}}\)}{{\KBB Tightened Big-M constants for synchronized-reserve bid-acceptance constraints.}}
\nomenclature[D]{\({\KBB M^{dir}}\)}{{\KBB Big-M constant used in the energy-price sign-regime constraints.}}
\nomenclature[D]{\(\pi_s\)}{Probability (weight) of scenario \(s\).}
\nomenclature[D]{\(\alpha_s\)}{Scenario scaling (perturbation) factor applied to base-case prices for scenario \(s\).}
\nomenclature[D]{\(p_{ch}(t)\)}{Hour-dependent probability of applying a sign change to the energy price at hour \(t\) during scenario generation.}
\nomenclature[D]{\({\KBB N_{\mathrm{gen}}}\)}{{\KBB Number of generated stochastic price scenarios before reduction.}}
\nomenclature[D]{\({\KBB N_s}\)}{{\KBB Number of representative scenarios after fast-forward reduction.}}
\nomenclature[D]{\({\KBB \sigma_{en},\,\sigma_{syn},\,\sigma_{\rho}}\)}{{\KBB Noise-scale parameters for energy price, synchronized-reserve price, and reserve-utilization probability scenario generation.}}
\nomenclature[D]{\(n^{\mathrm{bid}}\)}{Number of discrete bid segments per product and per hour.}
\nomenclature[D]{\(C_0\)}{Initial state of charge at the start of the optimization horizon. (MWh)}
\nomenclature[D]{\(\underline{C},\,\bar{C}\)}{Minimum and maximum allowable SoC (SoC bounds). (MWh)} 
\nomenclature[D]{\(\underline{h},\,\bar{h}\)}{Minimum and maximum hydraulic head.}
\nomenclature[D]{\(a^{ch}_i,c^{ch}_i,d^{ch}_i,e^{ch}_i\)}{Linear coefficients that define upper/lower bounding lines for \(\mathcal{F}^{ch}(h)\).}
\nomenclature[D]{\(a^{ds}_i,c^{ds}_i,d^{ds}_i,e^{ds}_i\)}{Linear coefficients that define upper/lower bounding lines for \(\mathcal{F}^{ds}(h)\).}


\section{Introduction}\label{sec:IN}
The increasing penetration of intermittent renewable energy sources, such as wind and solar photovoltaics, has substantially increased the variability of net load profiles, thereby posing significant operational challenges for modern power systems. Owing to their stochastic nature and limited controllability, these resources challenge conventional balancing mechanisms, intensifying the demand for system flexibility to sustain frequency stability, ramping adequacy, and reliable operation \cite{tian2020risk, RefA, RefB}. 
In particular, as renewable penetration increases to meet decarbonization targets, securing sufficient system flexibility at both operational and market levels has become indispensable \cite{RefC-1, RefC-2}.

Large-scale energy storage systems (ESSs) have emerged as key solutions to address these flexibility challenges. Among these, pumped storage hydropower (PSH) has long served as the backbone of grid-scale storage owing to its high efficiency, large capacity, and long-duration storage capability. It currently accounts for more than 90\% of global utility-scale energy storage capacity \cite{ren212022}. Moreover, PSH has been recognized as an essential technology to support renewable energy integration and maintain system reliability \cite{denholm2010role}. However, the majority of existing PSH units are fixed-speed, which face limitations such as the inability to continuously modulate pumping power, provide ancillary services during pumping mode, or adequately respond to market price signals \cite{vargas2017economic, RefD}.

Variable speed pumped storage hydropower (VS-PSH) has thus emerged as a promising alternative. By decoupling the turbine and generator rotational speeds, VS-PSH units enable continuous power modulation under varying head conditions and can even provide ancillary services while operating in pumping mode \cite{filipe2019optimal, RefE-1, RefE-2}. Recent studies have demonstrated substantial advances in the modeling, control, and operational optimization of VS-PSH, including detailed nonlinear representations and efficiency and transient-aware plant-level frameworks \cite{NEW-A, NEW-B, NEW-C}. 
{\KB More recently, dynamic efficiency formulations that capture the nonlinear relationship between hydraulic head, flow rate, and pump-turbine efficiency through piecewise linear interpolation techniques have also been applied within coordinated scheduling frameworks for pumped storage hydropower \cite{Reviewer3}.}
The deployment of VS-PSH has been expanding internationally, with projects reported in Asia and Europe, including Korea, Japan, Switzerland, and Austria \cite{hydropower2025world}. 
{\KB More recent reviews further document ongoing large-scale VS-PSH projects in China, such as Fengning Phase II, Taian Phase II, Langjian, and Zhongdong, as well as European projects including Frades II in Portugal and Nant de Drance in Switzerland \cite{Reviewer1-7_NEW1}.}
In the United States, interest in new VS-PSH developments has been growing, and these projects are increasingly regarded as a key option to enhance grid flexibility and support renewable integration \cite{martinez2023us}. Collectively, these developments underscore the rising recognition of VS-PSH as a viable long-duration storage solution across diverse power systems.

Extensive studies have been conducted on market participation and scheduling strategies for storage resources, including market-oriented operation and short-term scheduling frameworks for VS-PSH \cite{NEW-D, NEW-E}. Existing studies have introduced stochastic scheduling strategies for PSH to manage market uncertainties \cite{RefM}, and subsequent works have assessed the economic viability of retrofitting existing facilities with variable-speed technology \cite{RefN}. 
{\KB Recent operational evaluations further quantify the flexibility benefits of VS-PSH in renewable-rich energy systems, demonstrating measurable improvements in system flexibility, renewable accommodation, and reduction of unit start-up and shutdown operations compared with fixed-speed units \cite{Reviewer1-7_NEW2}.}
Some studies have demonstrated the profitability of VS-PSH participation in day-ahead energy and reserve markets \cite{filipe2019optimal}. More recent contributions have developed stochastic formulations for joint energy and reserve scheduling of energy-limited storage resources in day-ahead markets \cite{tang2020reserve}. Security-constrained reserve scheduling for pumped storage hydropower under uncertainty has also been investigated in ISO day-ahead markets \cite{liu2021secured}. Nevertheless, simplified single-segment bidding representations for PSH have been adopted in market-clearing formulations \cite{RefG-1}. Aggregated offer structures have similarly been used in stochastic market participation models \cite{RefG-2}. Market-oriented scheduling frameworks with limited bid segmentation have also been reported for PSH participation \cite{RefH-1}. Such assumptions restrict the adequate representation of marginal value in practical storage market implementations \cite{RefH-2}.

Overall, these approaches fall short of reflecting the complex operational characteristics of VS-PSH and the granularity required for market compatibility. VS-PSH exhibits head-dependent nonlinear output, discrete pumping–generation mode transitions, and requires energy continuity constraints. Variants of these constraints have been modeled in look-ahead dispatch frameworks from the system operator perspective for conventional fixed-speed PSH \cite{RefI}, but this line of research remains operator-centric and does not provide guidance on how VS-PSH owners can strategically participate in electricity markets. Several studies have attempted to investigate VS-PSH market participation under physical constraints \cite{RefJ}, but they often result in dispatch schedules rather than market-compatible bidding curves. Furthermore, restricting storage participation to single-segment bidding structures fundamentally limits the accurate reflection of marginal costs in actual market settings \cite{RefK}. Therefore, this study proposes an integrated framework that (i) explicitly accounts for physical characteristics of VS-PSH, (ii) simultaneously co-optimizes offers for energy and ancillary service, and (iii) generates multi-segment bidding curves aligned with market rules.

\begin{table*}[t] 
\centering
\caption{Comparison of PSH Scheduling Approaches under Market Considerations}
\label{tab:lit_review_compact}

\setlength{\tabcolsep}{5pt} 
\renewcommand{\arraystretch}{1.15}
\footnotesize

\begin{tabularx}{\textwidth}{%
>{\centering\arraybackslash}m{1.25cm}
>{\centering\arraybackslash}m{1.30cm}
>{\centering\arraybackslash}m{1.25cm}
>{\centering\arraybackslash}m{1.25cm}
>{\centering\arraybackslash}m{1.25cm}
>{\centering\arraybackslash}m{1.5cm}
X
}

\toprule
\textbf{Ref.} & 
\textbf{Type}$^{(a)}$ & 
\textbf{Multi Segment} & 
\textbf{Head Dependent} & 
\textbf{Energy - AS} & 
\textbf{Stochastic Uncertainty} & 
\textbf{Modeling Approach} \\

\midrule

\cite{NEW-D} & VS-PSH & \checkmark & \checkmark$^{(c)}$ & \xmark & \xmark & MINLP + robust opt. \\

\cite{NEW-E} & VS-PSH & \xmark & \xmark & \xmark & \checkmark & Multi-objective MILP \\

\cite{RefM} & VS-PSH & \checkmark & \xmark & \xmark& \xmark & MILP \\

\cite{tang2020reserve} & ES & \xmark & \xmark & \checkmark & \checkmark & Stochastic UC \\

\cite{liu2021secured} & FS-PSH & \xmark & \xmark & \checkmark & \checkmark & CC UC \\

\cite{RefG-1} & ES & \xmark & \xmark & \xmark & \checkmark & Bilevel MILP \\

\cite{RefG-2} & BS & \xmark & \xmark & \xmark & \xmark 
& Holistic simulation \\

\cite{RefI} & FS-PSH & \checkmark & \checkmark$^{(d)}$ & \xmark & \xmark  & Deterministic MILP \\

\cite{RefK} & Generic$^{(b)}$ & \xmark & \xmark & \xmark & \checkmark  & Stochastic programming \\

\addlinespace[2pt]
\textbf{This work} & VS-PSH  & \checkmark & \checkmark & \checkmark & \checkmark & Stochastic MILP \\
\bottomrule
\end{tabularx}

\vspace{2mm}
\begin{minipage}{\linewidth}
\raggedright
\scriptsize
\textit{Abbreviations} – AS: Ancillary Services, VS-PSH: Variable Speed Pumped Storage Hydropower, FS-PSH: Fixed Speed Pumped Storage Hydropower, ES: Energy Storage, BS: Battery Storage, MINLP: Mixed-Integer Nonlinear Programming, MILP: Mixed-Integer Linear Programming, UC: Unit Commitment, CC: Chance Constrained \\
$^{(a)}$ Type classifies studies based on whether PSH-specific physical constraints are explicitly modeled.
\\
$^{(b)}$ Generic storage model not specific to PSH technology.\\
$^{(c),}$ $^{(d)}$ Incorporates head-dependent efficiency under deterministic settings.

\end{minipage}
\end{table*}

Table~\ref{tab:lit_review_compact} compares recent approaches on market-oriented scheduling and bidding representations for PSH. While some prior studies incorporate either physical constraints or stochastic modeling, none to date provide a comprehensive solution that combines all these aspects in a single model. By contrast, our proposed model uniquely integrates multi-segment bidding and detailed VS-PSH physical constraints within a market-compatible stochastic MILP framework.

{\KB To fill these gaps, this study proposes a day-ahead optimal scheduling framework for VS-PSH. The market relevance of this work is further motivated by the declining effective load-carrying capability (ELCC) recently reported for long-duration energy storage resources in markets such as PJM \cite{RefO-1, RefO-2}. As capacity-credit-based remuneration contracts under this trend, energy arbitrage and ancillary service revenues become an increasingly important share of the total revenue stream available to VS-PSH owners. Accurately representing marginal cost through multi-segment offers therefore becomes more economically consequential, providing a direct motivation for the multi-segment, multi-product bidding framework developed in this paper. The model is formulated as a mixed-integer linear program (MILP) that simultaneously co-optimizes offers for energy and ancillary services while explicitly incorporating the physical constraints of VS-PSH.}
Crucially, it generates multi-segment bidding curves for each hourly market product, in contrast to existing methods that mainly adopt single-segment bidding structures \cite{tang2020reserve}, \cite{liu2021secured}, while remaining consistent with market formulations that allow multi-segment bids \cite{RefL}. 
{\KB The regulatory basis for storage resources—including pumped storage hydropower—to submit segmented offers in U.S. wholesale electricity markets is established by FERC Order No. 841 \cite{Reviewer1-7_NEW3}, which required ISOs and RTOs to develop participation models enabling electric storage resources to compete in energy, capacity, and ancillary service markets.}
This multi-segment structure allows VS-PSH to better reflect its marginal costs and operational flexibility in market settings, ultimately improving both profitability and system support. Unlike existing studies that mainly focus on system operation, this study presents a market-oriented scheduling framework from the perspective of market participants.

The main contributions of this study are summarized as follows:
\begin{itemize}
\item Development of an integrated day-ahead co-optimization model for VS-PSH offers across energy and ancillary services.

\item Formulation of multi-segment hourly bidding curves for each product, enhancing market compatibility.

\item Explicit incorporation of VS-PSH physical constraints, including head-dependent capability limits, discrete pumping and generating mode transitions, and energy continuity constraints, within the market model.

\item A case study validating the practical applicability of the proposed multi-segment, multi-product bidding framework under realistic operational constraints.
\end{itemize}

The remainder of this paper is organized as follows. Section~\ref{sec:MODEL} presents the modeling framework for VS-PSH scheduling in day-ahead energy and ancillary service markets. Section~\ref{sec:SOL} describes the solution implementation, including scenario generation and stochastic optimization. Section~\ref{sec:NT} discusses numerical case studies that validate the proposed framework under varying market conditions. Finally, Section~\ref{sec:CON} presents the conclusion with a summary of key findings and directions for future research.


\section{Modeling the Variable-Speed Pumped-Storage Hydropower Scheduling Framework}\label{sec:MODEL}
\subsection{Energy market modeling}
We define binary variables $x_t^{\mathrm{ch}},\;x_t^{\mathrm{ds}},\;x_t^{\mathrm{con}},\;x_t^{non}$ to represent the operational mode of the VS-PSH unit at each time $t$. The modes are charging, discharging, continuous, and non-operational, respectively. The selection is governed by the following constraint:
\begin{align}
x_t^{\mathrm{ch}} + x_t^{\mathrm{ds}} + x_t^{\mathrm{con}} + x_t^{\mathrm{non}} = 1. \label{eq:mode}
\end{align}
Constraint \eqref{eq:mode} enforces the mode-selection constraint, ensuring that for any given time $t$, the VS-PSH unit operates in exactly one of the four possible and mutually exclusive modes.

\vphantom{a}
\noindent
Suppose that the VS-PSH operator is interested in submitting a bid for a range of dis(charging) values. We discretize each (dis)charging range by choosing $n^{\text{bid}}$ bidding values
$\{\hat{b}^{ch}_{i}\}_{i=1}^{n^{\text{bid}}}$ and $\{\hat{b}^{ds}_{i}\}_{i=1}^{n^{\text{bid}}}$.
The bidding prices for these discrete power levels are denoted by $\{\hat p^{ch}_{i,t}\}_{i=1}^{n^{\text{bid}}}$ and $\{\hat p^{ds}_{i,t}\}_{i=1}^{n^{\text{bid}}}$.
To ensure continuity between charging and discharging, charging quantities are represented as negative, i.e., $b^{ch}_{t,s}\leq 0$ and $b^{ch}_{t,s}\in [-\bar b^{ch}_{t,s},\,-\underline b^{ch}_{t,s}]$. Moreover, we assume $\hat p^{ch}_{i,t}<0$ for charging prices.

\begin{figure}[t]
    \centering
    \includegraphics[width=\linewidth]{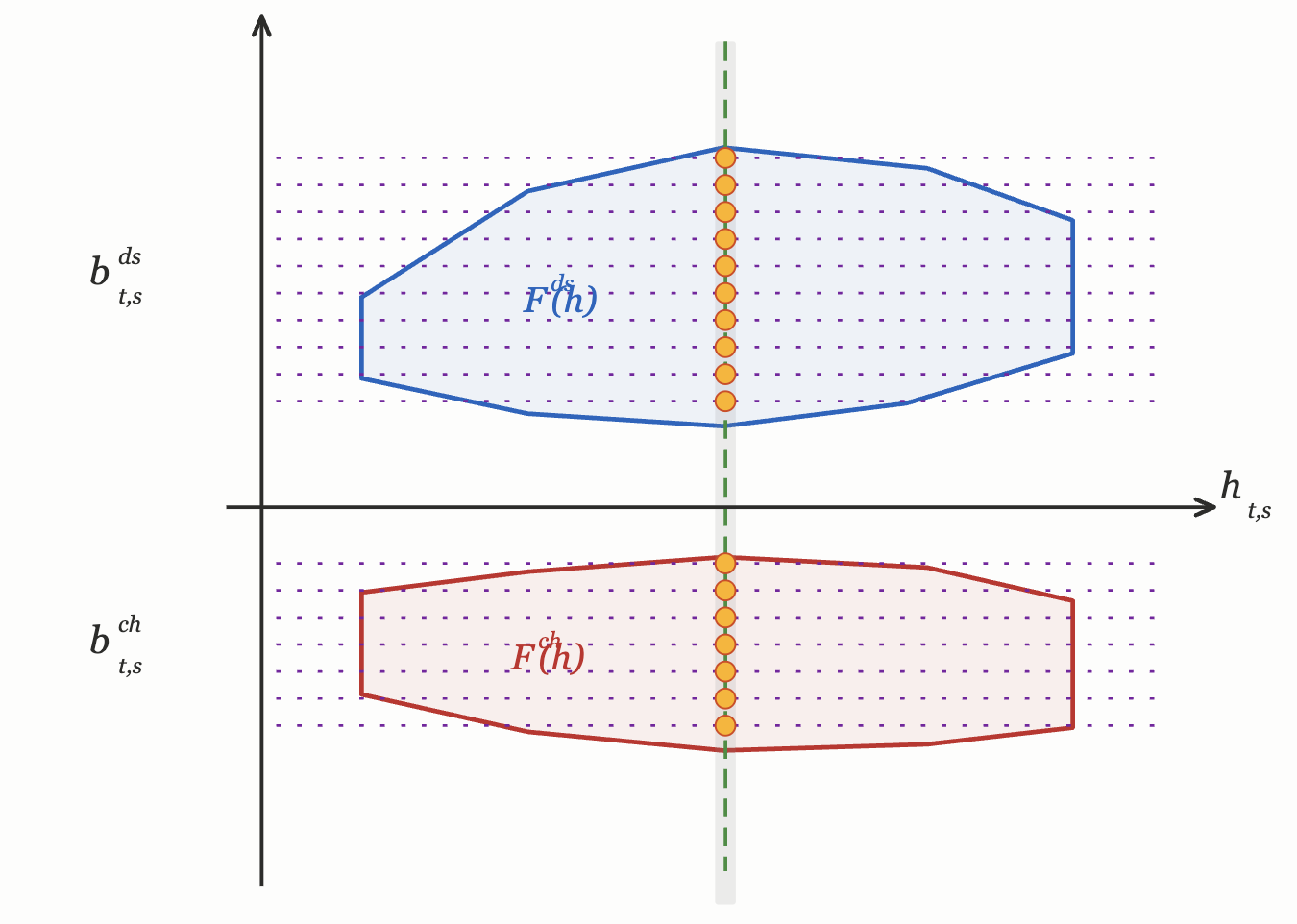}  
    \caption{\small Head-dependent feasible sets for VS-PSH charging and discharging. The blue polygon represents the border of the feasible region for discharging power, and the red polygon represents the border of the feasible region for charging power. Purple dotted lines represent the complete set of discretized power levels, and the yellow dots represent a specific instance of the feasible charge/discharge bid levels.}
    \label{fig:head}
\end{figure}

\vphantom{a}
\noindent
Additionally, we considered the head-dependent nature of pumped storage operations, as illustrated in Fig.~\ref{fig:head}. As shown in the figure, the (dis)charging power is constrained based on the head level. Once the head level is determined, the maximum and minimum (dis)charging powers, along with the corresponding available outputs, are established. These available outputs represent the power that the VS-PSH can bid. Mathematically, we define the head-to-charge and head-to-discharge feasible sets as follows:
\begin{equation}
\mathcal{F}^{ch}(h) = \left\{\hat{b}\;\middle|\;
\begin{aligned}
& \hat{b} \le a^{ch}_i h + c^{ch}_i, && \forall\, i \in \mathcal{I}^{ch}_1, \\
& \hat{b} \ge d^{ch}_i h + e^{ch}_i, && \forall\, i \in \mathcal{I}^{ch}_2
\end{aligned}
\right\}. \label{eq:f_ch}
\end{equation}
\begin{equation}
\mathcal{F}^{ds}(h) = \left\{\hat{b}\;\middle|\;
\begin{aligned}
& \hat{b} \le a^{ds}_i h + c^{ds}_i, && \forall\, i \in \mathcal{I}^{ds}_1, \\
& \hat{b} \ge d^{ds}_i h + e^{ds}_i, && \forall\, i \in \mathcal{I}^{ds}_2
\end{aligned}
\right\}. \label{eq:f_ds}
\end{equation}
The sets \eqref{eq:f_ch} and \eqref{eq:f_ds} mathematically describe the head-dependent operational limits of the VS-PSH. $\mathcal{F}^{ch}(h)$ defines the feasible range of charging power, and $\mathcal{F}^{ds}(h)$ defines the feasible range of discharging power, both as a function of the hydraulic head $h$. These ranges are represented by a set of linear inequalities. The variable speed constraint can be expressed as
\begin{align}
    &{b}^{ch}_{t,s} \in \mathcal{F}^{ch}(h_{t,s}), \quad {b}^{ds}_{t,s} + b^{syn}_{t,s} \in \mathcal{F}^{ds}(h_{t,s}) \label{eq:b_in_f}
\end{align}
These constraints enforce that the scheduled charging power $b^{ch}_{t,s}$ and discharging power $b^{ds}_{t,s}$ with synchronized reserve $b^{syn}_{t,s}$ for each time $t$ and scenario $s$ must respect the physical operating limits dictated by the current head $h_{t,s}$.

\vphantom{a}
\noindent
In addition, based on the mode selection, the activation of a bid is constrained.
\begin{subequations}
\label{eq:mode_link}
\begin{align}
    &z^{ch}_{i,t,s} \leq x^{con}_t+x^{ch}_t,\\
    &z^{ds}_{i,t,s} \leq x^{con}_t+x^{ds}_t.
\end{align}
\end{subequations}
Constraints \eqref{eq:mode_link} link the bidding acceptance variables ($z^{ch}_{i,t,s}, z^{ds}_{i,t,s}$) to the operational mode. A charging bid can only be selected if the unit is in the charging or continuous mode. Similarly, a discharging bid can only be selected if the unit is in the discharging or continuous mode. The continuous mode allows the unit to bid in both charging and discharging simultaneously.

\vphantom{a}
\noindent
Once the mode and bidding price for each charging or discharging segment are determined, the bid is accepted. This process involves uncertainty and is represented using a scenario-based approach.
For each scenario $s$, let $\delta^{en}_{t,s}$ denote the cleared price at time $t$. If $\delta^{en}_{t,s}>0$ the discharging bid is accepted, whereas if $\delta^{en}_{t,s}<0$ the charging bid is accepted.
Next, we introduce binary variables $z^{ch}_{i,t,s}$ and $z^{ds}_{i,t,s}$, $i=1,\dots,n^{\text{bid}}$, to enforce that exactly one segment (either charging or discharging) is selected.
\begin{align}
    \sum_{i=1}^{n^{\text{bid}}} z^{ch}_{i,t,s} \;+\; \sum_{i=1}^{n^{\text{bid}}} z^{ds}_{i,t,s} \;=\; 1-x^{non}_{t}. \label{eq:one_bid_selected}
\end{align}
Equation \eqref{eq:one_bid_selected} ensures that if the VS-PSH unit is operational (i.e., not in the `non' mode, $x^{non}_t = 0$), exactly one of the discrete bid segments (either for charging or discharging) is selected and cleared in the market for each scenario $s$ at time $t$.

\vphantom{a}
\noindent
The following set of constraints models the market clearing logic based on bid prices and the market clearing price $\delta^{en}_{t,s}$.
\begin{subequations}
\label{eq:market_clearing_energy}
\begin{align}
    &\delta^{en}_{t,s} (0.5-\gamma^{ds}_{t,s}) < 0,\label{eq:mcl_energy_a}\\
    &\hat{p}^{ch}_{i,t} \geq \delta^{en}_{t,s} - {\KBB M^{ch}} (1-\hat{z}^{ch}_{i,t,s}),\label{eq:mcl_energy_b}\\
    &\hat{p}^{ds}_{i,t} \leq \delta^{en}_{t,s} + {\KBB M^{ds}} (1-\hat{z}^{ds}_{i,t,s}),\label{eq:mcl_energy_c}\\
    &\hat{z}^{ch}_{i,t,s} \leq 1-\gamma^{ds}_{t,s},\label{eq:mcl_energy_d}\\
    &\hat{z}^{ds}_{i,t,s} \leq \gamma^{ds}_{t,s},\label{eq:mcl_energy_e}\\
    &{z}^{ch}_{i,t,s} \leq {z}^{ch}_{j,t,s}, \quad \forall j < i,\label{eq:mcl_energy_f}\\
    &{z}^{ds}_{i,t,s} \leq {z}^{ds}_{j,t,s}, \quad \forall i < j,\label{eq:mcl_energy_g}\\
    &z^{ch}_{i,t,s} \leq \hat{z}^{ch}_{i,t,s},\label{eq:mcl_energy_h}\\
    &z^{ds}_{i,t,s} \leq \hat{z}^{ds}_{i,t,s}\label{eq:mcl_energy_i}
\end{align}
\end{subequations}
The block of constraints in \eqref{eq:market_clearing_energy} defines the bid acceptance conditions. Constraint \eqref{eq:mcl_energy_a} uses a binary variable $\gamma^{ds}_{t,s}$ to determine if the clearing price is for charging (negative) or discharging (positive). Constraints \eqref{eq:mcl_energy_b} and \eqref{eq:mcl_energy_c} are Big-M constraints that ensure a bid is accepted only if its price is competitive against the market price $\delta^{en}_{t,s}$. Specifically, a charging bid price $\hat{p}^{ch}$ must be at or above the clearing price, and a discharging bid price $\hat{p}^{ds}$ must be at or below it. Constraints \eqref{eq:mcl_energy_d} and \eqref{eq:mcl_energy_e} link the feasible bid acceptance variables $\hat{z}$ to the price sign variable $\gamma^{ds}_{t,s}$. Constraints \eqref{eq:mcl_energy_f} and \eqref{eq:mcl_energy_g} ensure that the bid with the largest (dis)charging power is finally accepted. Finally, \eqref{eq:mcl_energy_h} and \eqref{eq:mcl_energy_i} link the feasible bid acceptance variables $\hat{z}$ to the final bid selection variables $z$.
{\KBB To tighten the MILP relaxation and improve numerical robustness, we use scenario-dependent tightened constants rather than an arbitrarily large scalar. Let $\overline{\Delta}^{en}=\max_{t,s}\delta^{en}_{t,s}$, $\underline{\Delta}^{en}=\min_{t,s}\delta^{en}_{t,s}$, and define bid-price box bounds $\hat p^{ch}_{i,t}\in[\underline P^{ch},0]$, $\hat p^{ds}_{i,t}\in[0,\overline P^{ds}]$, $\hat p^{syn}_{i,t}\in[0,\overline P^{syn}]$. We set}
\begin{align}
{\KBB M^{ch}}&{\KBB=\overline{\Delta}^{en}-\underline P^{ch},\quad
M^{ds}=\overline P^{ds}-\underline{\Delta}^{en}.}
\end{align}
{\KBB In implementation, the bounds are taken from the generated scenario set each run, yielding the smallest valid Big-M values that preserve the bid-acceptance logic.}

\vphantom{a}
\noindent
The accepted (dis)charging amount is the sum of the accepted charging and discharging contributions,
\begin{align}
    b_{t,s} &= b^{ch}_{t,s} + b^{ds}_{t,s}, \label{eq:b_total}
\end{align}
where the individual components are determined by the accepted bid quantities:
\begin{align}
    b^{ch}_{t,s} = \sum_{i=1}^{n^{\text{bid}}} \hat{b}^{ch}_{i}\, z^{ch}_{i,t,s}, \quad
    b^{ds}_{t,s} = \sum_{i=1}^{n^{\text{bid}}} \hat{b}^{ds}_{i}\, z^{ds}_{i,t,s}. \label{eq:b_acc}
\end{align}
As exactly one $z$ variable is non-zero (equal to 1) for an operational unit, these equations assign the power quantity $\hat{b}^{ch}_{i}$ or $\hat{b}^{ds}_{i}$ corresponding to the single accepted bid.

\vphantom{a}
\noindent
Finally, the head is related to the state of charge (SoC) of the reservoir:
\begin{subequations}
\begin{align}
    & h_{t, s} \geq \left( \frac{C_{t, s} - \underline{C}}{\bar{C} - \underline{C}} \right) \cdot (\bar{h} - \underline{h}) + \underline{h} \label{eq:head_soc_relation_a} \\
& h_{t, s} \leq \left( \frac{C_{t, s} - \underline{C}}{\bar{C} - \underline{C}} \right) \cdot (\bar{h} - \underline{h}) + \underline{h} \label{eq:head_soc_relation_b}
\end{align}
\end{subequations}
The inequalities \eqref{eq:head_soc_relation_a} and \eqref{eq:head_soc_relation_b} represent a linearized relationship between the SoC $C_{t,s}$ and hydraulic head $h_{t,s}$. They constrain the head to be within a narrow band around a value that scales linearly with the SoC of the reservoir relative to its minimum ($\underline{C}$) and maximum ($\bar{C}$) levels.

\begin{remark}[Multi-segment bidding and market-compatibility]\label{rmk:multiseg}
The discretization of charging and discharging ranges into $n^{\mathrm{bid}}$ segments (see \eqref{eq:b_acc} and Fig.~\ref{fig:head}) is not merely a numerical convenience but a deliberate modeling choice to produce market-compatible, piecewise-constant bidding curves that respect head-dependent operational bounds $\mathcal{F}^{ch}(h)$ and $\mathcal{F}^{ds}(h)$ in \eqref{eq:f_ch}--\eqref{eq:f_ds}. By representing offers as a finite set of ordered bid segments and enforcing logical acceptance constraints (e.g., \eqref{eq:mcl_energy_f}--\eqref{eq:mcl_energy_i}), the framework maps physical capability directly into the market bid format used by many ISOs. This formulation is one of the key contributions of this paper by explicitly combining head-dependent feasibility with multi-segment, market-ready bid curves for VS-PSH, which preserves physical feasibility at the bidding stage and formulates a practical market bidding framework.
\end{remark}

\subsection{Ancillary-service market modeling}
In addition to the energy market, we consider that VS-PSH can participate in ancillary-service markets, especially the synchronized reserve markets.
Similar to the energy market, we use $n^{\text{bid}}$ discrete bid levels and denote the cleared prices by $\delta^{syn}_{t,s}$. The main difference is that reserve offers consider only discharging, i.e., generation to the system. The offered reserve capacity must be physically feasible by following \eqref{eq:b_acc}.


\vphantom{a}
\noindent
The bidding submission and clearing logic for the synchronized reserve market is analogous to the energy market:
\begin{subequations}
\label{eq:market_clearing_as}
\begin{align}
    &z^{syn}_{j,t,s} \leq x^{con}_{t} + x^{ds}_{t},\label{eq:mcl_as_a}\\
    &\sum_{j=1}^{n^{\text{bid}}} z^{syn}_{j,t,s} \;=\; 1 - x^{ch}_{t} - x^{non}_{t},\label{eq:mcl_as_b}\\
    &\delta^{syn}_{t,s} (0.5-\gamma^{syn}_{t,s}) < 0,\label{eq:mcl_as_c}\\
    &\hat{p}^{syn}_{i,t} \leq \delta^{syn}_{t,s} + {\KBB M^{syn}} (1-\hat{z}^{syn}_{i,t,s}),\label{eq:mcl_as_d}\\
    &\hat{z}^{syn}_{i,t,s} \leq \gamma^{syn}_{t,s},\label{eq:mcl_as_e}\\
    &{z}^{syn}_{i,t,s} \leq {z}^{syn}_{j,t,s}, \quad \forall i < j,\label{eq:mcl_as_f}\\
    &z^{syn}_{i,t,s} \leq \hat{z}^{syn}_{i,t,s}\label{eq:mcl_as_g}
\end{align}
\end{subequations}
In \eqref{eq:market_clearing_as}, constraint \eqref{eq:mcl_as_a} restricts reserve provision to the discharging and continuous modes. Equation \eqref{eq:mcl_as_b} ensures that if the unit is not charging or non-operational, exactly one reserve bid is selected. The remaining constraints \eqref{eq:mcl_as_c}-\eqref{eq:mcl_as_g} represent the market clearing logic, which demonstrates the energy market formulation: a reserve bid is accepted if its offer price $\hat{p}^{syn}$ is less than or equal to the clearing price for reserves $\delta^{syn}_{t,s}$.
{\KBB For \eqref{eq:mcl_as_d}, we use the tightened constant $M^{syn}=\overline P^{syn}$ so that the reserve-price acceptance inequality is relaxed only by the minimum valid amount when $\hat z^{syn}_{i,t,s}=0$.}

\vphantom{a}
\noindent
The awarded reserve capacity is determined by the accepted bid:
\begin{align}
    b^{syn}_{t,s} &= \sum_{i=1}^{n^{\text{bid}}} \hat{b}^{ds}_{i}\, z^{syn}_{i,t,s}. \label{eq:b_syn_acc}
\end{align}
Equation \eqref{eq:b_syn_acc} calculates the total cleared synchronized reserve capacity $b^{syn}_{t,s}$ by summing the quantities of the accepted bids. As only one $z^{syn}_{i,t,s}$ can assume the value 1, this effectively selects the power quantity from the single cleared bid.

\subsection{Integration of energy and ancillary service markets}
As the energy and ancillary-service markets are co-optimized, accepted (dis)charging and reserve capacities must satisfy the operational bounds. SoC management must be essential in the operation of VS-PSH, as improper management can result in a violation of the accepted bid and subsequently lead to penalties.
By representing $\rho^{syn}_{t,s}$ as the ratio that the reserve is used in the operation, we have
{\KBB Note that $\rho^{syn}_{t,s}$ is treated as a scenario-specific stochastic input---not a deterministic ex-ante parameter---estimated from historical synchronized-reserve prices as described in Section~\ref{sec:NT}.}
\begin{subequations}
\label{eq:integration_bounds}
\begin{align}
     &-\bar{b}^{ch}_{t,s} \leq b^{ch}_{t,s} \leq -\underline{b}^{ch}_{t,s},\label{eq:integration_a}\\
     &\underline{b}^{ds}_{t,s} \leq b^{ds}_{t,s} + b^{syn}_{t,s} \rho^{syn}_{t,s} \leq \bar{b}^{ds}_{t,s}.\label{eq:integration_b}
\end{align}
\end{subequations}
The set of constraints in \eqref{eq:integration_bounds} integrate energy and reserve market commitments. Constraint \eqref{eq:integration_a} bounds the scheduled charging power within its physical limits. Constraint \eqref{eq:integration_b} is the key coupling constraint, ensuring that the total power output---comprising the energy market dispatch ($b^{ds}_{t,s}$) and deployed portion of the synchronized reserve commitment ($\rho^{syn}_{t,s} b^{syn}_{t,s}$)---do not violate the physical discharging limits of the unit.

\vphantom{a}
\noindent
Let $C_{t,s}$ denote the SoC at time $t$ under scenario $s$, and let the (dis)charging efficiency satisfy $0<\eta<1$. Then, the SoC update is more consistently expressed as
\begin{subequations}
\label{eq:soc}
\begin{align}
    C_{t+1,s} &= C_{t,s} - \frac{1}{\eta}\,\big(b^{ds}_{t,s} + b^{syn}_{t,s}\rho^{syn}_{t,s}\big) - \eta\, b^{ch}_{t,s},\label{eq:soc_a}\\
    \underline C &\leq C_{t,s} \leq \bar C.\label{eq:soc_b}
\end{align}
\end{subequations}
SoC dynamics are described by \eqref{eq:soc}. Equation \eqref{eq:soc_a} models the energy balance of the reservoir: the SoC at time $t+1$ is calculated from the SoC at time $t$, accounting for energy withdrawn for discharging (in both energy and deployed ancillary service markets) and energy added from charging. 
{\KB In Equation \eqref{eq:soc_a} $\eta$ denotes the one-way efficiency factor. Charging energy is scaled by $\eta$ to reflect conversion losses during pumping, and discharging energy is scaled by $1/\eta$ to reflect conversion losses during generation. Accordingly, the round-trip efficiency of a complete charge--discharge cycle is $\eta^2$. This single parameter represents a lumped approximation that aggregates converter, mechanical, and hydraulic losses.} 
Equation \eqref{eq:soc_b} ensures that the energy level of the reservoir remains within its operational minimum ($\underline C$) and maximum ($\bar C$) bounds at all times.
Finally, the head $h_{t,s}$ is updated to $h_{t+1,s}$ based on the linear relationship between SoC and head position.

\vphantom{a}
\noindent
Along with these dynamics, the objective is to maximize the profit of the pumped-storage owner over the optimization horizon $\mathcal{T}$ in the day-ahead market:
\begin{align}
    \max \; \sum_{t \in \mathcal{T}} \sum_{s \in \mathcal{S}} \pi_s \Big[\,
      &\delta^{en}_{t,s}\, (b^{ds}_{t,s}-b^{ch}_{t,s})
      + \delta^{syn}_{t,s}\, b^{syn}_{t,s}\Big]. \label{eq:obj}
\end{align}
The profit of each period is the sum of revenues from the energy market (revenue from selling power $b^{ds}_{t,s}$ minus cost of buying power $b^{ch}_{t,s}$, at price $\delta^{en}_{t,s}$) and the ancillary service market (revenue from providing reserve capacity $b^{syn}_{t,s}$ at price $\delta^{syn}_{t,s}$). The profit of each scenario is weighted by its probability $\pi_s$.

\begin{remark}[Simultaneous co-optimization of energy and synchronized reserves]\label{rmk:coopt}
The co-optimization of energy and ancillary services is enforced by the coupling constraints in \eqref{eq:integration_bounds} and the SoC dynamics in \eqref{eq:soc}. Modeling reserve commitments as cleared bid segments ($b^{syn}_{t,s}$) together with a deployment ratio $\rho^{syn}_{t,s}$ explicitly captures the operational trade-off between energy dispatch and reserve availability: cleared reserve must be physically feasible under the reservoir head and cannot be treated independently of energy schedules. This integrated treatment (i) prevents infeasible ex-post deployments that would violate head-dependent limits, (ii) internalizes the opportunity cost of reserving capacity for reserves instead of energy, and (iii) permits evaluation of joint revenue streams within a single, market-compatible MILP. Integrating multi-segment bidding across both energy and synchronized-reserve products for a head-dependent VS-PSH model in a single co-optimization framework is novel and materially improves the model's practical relevance for market participation analysis.
\end{remark}

\section{Solution Implementation}
\label{sec:SOL}

{\KB Based on the problem formulation in Section~\ref{sec:MODEL}, we developed a computational framework to determine the optimal day-ahead scheduling strategy for a VS-PSH operator. This process involves two main stages: first, generating stochastic market scenarios and reducing them to a compact representative set, and second, solving a stochastic MILP to derive the optimal bidding curves and operational modes. The entire workflow is detailed in Algorithms~\ref{alg:scenario_generation} and \ref{alg:VS-PSH_optimization}.}

\begin{algorithm}[t!]
\SetAlgoLined
\caption{{\KB Stochastic Scenario Generation and Fast-Forward Scenario Reduction}}
\label{alg:scenario_generation}
\DontPrintSemicolon
{\bf Inputs:} Base profiles $\{\delta^{en,base}_{t},\delta^{syn,base}_{t},\rho^{syn,base}_{t}\}_{t\in\mathcal{T}}$, generated scenario count $N_{\mathrm{gen}}$, reduced scenario count $N_s$, AR(1) parameter $\phi$, noise scales $(\sigma_{en},\sigma_{syn},\sigma_{\rho})$, random seed.\;
\BlankLine
Generate standardized AR(1) noise trajectories $\{\epsilon_{t,j}\}_{t\in\mathcal{T},j=1,\ldots,N_{\mathrm{gen}}}$, where scenario $j=1$ is the baseline trajectory (all-zero noise).\;
\BlankLine
\For{$j \in \{1,\dots,N_{\mathrm{gen}}\}$}{
    Create generated energy scenarios:
    $\delta^{en}_{t,j} \leftarrow \delta^{en,base}_{t} + \sigma_{en}\epsilon_{t,j}$.\;
    Create generated synchronized-reserve price scenarios:
    $\delta^{syn}_{t,j} \leftarrow \max\left(0,\delta^{syn,base}_{t} + \sigma_{syn}\epsilon_{t,j}\right)$.\;
    Create generated reserve-acceptance probability scenarios:
    $\rho^{syn}_{t,j} \leftarrow \mathrm{clip}\!\left(\rho^{syn,base}_{t} + \sigma_{\rho}\epsilon_{t,j},0,1\right)$.\;
}
\BlankLine
Set initial probabilities: $\pi^{0}_j \leftarrow 1/N_{\mathrm{gen}}$ for all $j$.\;
\BlankLine
Compute pairwise scenario distances:
$d(j,k)=\left\|\delta^{en}_{:,j}-\delta^{en}_{:,k}\right\|_2$.\;
Initialize representative set: $\mathcal{R}\leftarrow\{1\}$ (always keep baseline).\;
\While{$|\mathcal{R}|<N_s$}{
    For each non-selected scenario $j$, compute
    $m_j=\min_{r\in\mathcal{R}} d(j,r)$.\;
    Select $j^\star=\arg\max_{j\notin\mathcal{R}} \pi^{0}_j m_j$ and update
    $\mathcal{R}\leftarrow\mathcal{R}\cup\{j^\star\}$.\;
}
\BlankLine
Assign each original scenario $j$ to its nearest representative in $\mathcal{R}$ and aggregate probabilities:
$\pi_r=\sum_{j\in\mathcal{N}(r)}\pi^{0}_j$.\;
\BlankLine
{\bf Return:} Reduced scenarios $\{\delta^{en}_{t,s},\delta^{syn}_{t,s},\rho^{syn}_{t,s}\}_{s=1}^{N_s}$ and reduced probabilities $\{\pi_s\}_{s=1}^{N_s}$.
\end{algorithm}

\begin{algorithm}[t]
\SetAlgoLined
\caption{VS-PSH scheduling and operation optimization}
\label{alg:VS-PSH_optimization}
\DontPrintSemicolon
{\bf Inputs:} Initial SoC $C_0$, reservoir parameters $(h_{\min}, h_{\max}, C_{\min}, C_{\max})$, efficiency $\eta$, head-dependent feasible sets $\mathcal{F}^{ch}(h)$, and $\mathcal{F}^{ds}(h)$.\;
\BlankLine
\tcp{Step 1: Generate price scenarios}
{\KB Use Algorithm~\ref{alg:scenario_generation} to generate $N_{\mathrm{gen}}$ stochastic scenarios and reduce them to $S$ representative scenarios, yielding $\{\delta^{en}_{t,s}\}_{t \in \mathcal{T}, s \in \mathcal{S}}$, $\{\delta^{syn}_{t,s}\}_{t \in \mathcal{T}, s \in \mathcal{S}}$, $\{\rho^{syn}_{t,s}\}_{t \in \mathcal{T}, s \in \mathcal{S}}$, and probabilities $\{\pi_s\}_{s \in \mathcal{S}}$.\;}
\BlankLine
\tcp{Step 2: Formulate MILP optimization problem}
{\bf Objective Function:} Maximize \eqref{eq:obj}\;
{\bf Subject to:} Constraints \eqref{eq:mode}-\eqref{eq:soc}\;
\BlankLine
\tcp{Step 3: Solve MILP problem}
Solve the mixed-integer linear programming problem using a commercial solver\;
\BlankLine
\tcp{Step 4: Extract optimal solutions}
Extract optimal mode selection variables: $\{x_t^{\mathrm{ch}}, x_t^{\mathrm{ds}}, x_t^{\mathrm{con}}, x_t^{\mathrm{non}}\}_{t \in \mathcal{T}}$\;
Extract optimal bidding prices:$\{\hat{p}^{\mathrm{ch}}_{i,t}, \hat{p}^{\mathrm{ds}}_{i,t}, \hat{p}^{\mathrm{syn}}_{i,t}\}_{i=1,\ldots,n^{\text{bid}}, t \in \mathcal{T}}$\;
Energy market charging prices: $\{\hat{p}^{\mathrm{ch}}_{i,t}\}_{i=1,\ldots,n^{\text{bid}}, t \in \mathcal{T}}$\;
Energy market discharging prices: $\{\hat{p}^{\mathrm{ds}}_{i,t}\}_{i=1,\ldots,n^{\text{bid}}, t \in \mathcal{T}}$\;
Synchronized reserve prices: $\{\hat{p}^{\mathrm{syn}}_{i,t}\}_{i=1,\ldots,n^{\text{bid}}, t \in \mathcal{T}}$\;
Extract optimal operational variables: $\{b_{t,s}, b^{ch}_{t,s}, b^{ds}_{t,s}, b^{syn}_{t,s}, C_{t,s}, h_{t,s}\}_{t \in \mathcal{T}, s \in \mathcal{S}}$\;
\BlankLine
{\bf Return:} Optimal mode selection $\{x_t^{\mathrm{ch}}, x_t^{\mathrm{ds}}, x_t^{\mathrm{con}}, x_t^{\mathrm{non}}\}_{t \in \mathcal{T}}$ and optimal bidding prices $\{\hat{p}^{\mathrm{ch}}_{i,t}, \hat{p}^{\mathrm{ds}}_{i,t}, \hat{p}^{\mathrm{syn}}_{i,t}\}_{i=1,\ldots,n^{\text{bid}}, t \in \mathcal{T}}$.
\end{algorithm}

{\KBB First, we generated market-price scenarios to capture future uncertainty using Algorithm~\ref{alg:scenario_generation}. The algorithm starts from one base profile, which represents the day-ahead forecasted baseline, and creates a larger stochastic set ($N_{\mathrm{gen}}=360$) using AR(1)-correlated noise for the energy price, synchronized reserve price, and reserve acceptance probability trajectories. Since day-ahead markets typically anticipate overarching daily trends (e.g., peak and off-peak hours) with reasonable accuracy, this approach intentionally preserves the underlying structural pattern while capturing realistic uncertainties in exact price magnitudes and localized volatility. In this study, the noise-scale parameters were set to $(\sigma_{en},\sigma_{syn},\sigma_{\rho})=(0.28,0.32,0.16)$ to reflect an enlarged uncertainty envelope. Next, a fast-forward reduction step is applied to select $N_s=5$ representative scenarios. The representative set is built by iteratively selecting scenarios that maximize probability-weighted distance from the already selected set, while preserving the baseline scenario. Finally, probabilities are reassigned by mapping each generated scenario to its nearest representative and aggregating the original uniform probabilities. The reduced scenario set and associated probabilities are then used in the stochastic MILP.}

{\KB Next, the VS-PSH scheduling and operation problem is solved as outlined in Algorithm~\ref{alg:VS-PSH_optimization}. This algorithm takes the reduced scenario set from Algorithm~\ref{alg:scenario_generation} and the VS-PSH physical parameters as inputs. The objective is to maximize the expected profit over all representative scenarios, subject to the full set of operational constraints, including mode selection, head-dependent capabilities, and energy continuity, as formulated in Section~\ref{sec:MODEL}. This stochastic optimization problem is solved using a commercial MILP solver.}

The key output of the optimization is the day-ahead scheduling strategy, which comprises decisions made before the realization of market prices. These are scenario-independent decisions, including the optimal hourly operating modes ($x_t^{\mathrm{ch}}, x_t^{\mathrm{ds}}, \dots$) and the multi-segment bidding prices for charging ($\hat{p}^{\mathrm{ch}}_{i,t}$), discharging ($\hat{p}^{\mathrm{ds}}_{i,t}$), and synchronized reserve ($\hat{p}^{\mathrm{syn}}_{i,t}$). 
Ultimately, the algorithm returns the optimal bidding prices and mode selections that the VS-PSH operator submits to the day-ahead market.


%

\section{Numerical Test} \label{sec:NT}

In this section, we present a numerical case study to validate the effectiveness of the proposed VS-PSH scheduling framework. The simulation is performed for a 24-h scheduling horizon in a day-ahead market, based on a representative set of scenarios. The optimization model is implemented in Python using the Gurobipy solver.
The results of the optimization provide a comprehensive day-ahead strategy, including hourly operational modes, multi-step bidding curves, and the expected financial performance.

\begin{remark}
The numerical experiments employ actual 2025 LMP and synchronized-reserve price series obtained from PJM \cite{pjm_dataminer}. Because publicly available datasets do not include unit-level bid-acceptance records (i.e., whether individual charging or discharging bids were cleared), acceptance of charging vs.\ discharging was modeled as a random draw to reflect market-clearing uncertainty. Similarly, synchronized-reserve acceptance probabilities were inferred from historical synchronized-reserve prices by scaling each observed hourly reserve price against a high-percentile reference level (the 81st-largest historical value) and capping the resulting ratio at 1.0. These hour-by-hour probabilities were sampled across a large set of randomly selected days from the 2025 dataset to generate scenario realizations of cleared energy and reserve outcomes.  
\end{remark}

{\KB In the revised scenario-generation pipeline, both energy prices and synchronized-reserve prices are treated as scenario-dependent stochastic inputs rather than keeping reserve prices fixed across scenarios. Specifically, Algorithm~\ref{alg:scenario_generation} perturbs $\delta^{en}_{t,s}$, $\delta^{syn}_{t,s}$, and $\rho^{syn}_{t,s}$ using correlated AR(1) disturbances, then reduces the generated set to representative scenarios. This preserves co-movement between energy and reserve conditions while keeping the stochastic MILP tractable.}

\subsection{Case Study \#1: Case \textit{HIGH}}
\begin{figure}[t]
    \centering
    \subfigure[]{\label{fig:da_price_high}\includegraphics[width=0.48\linewidth]{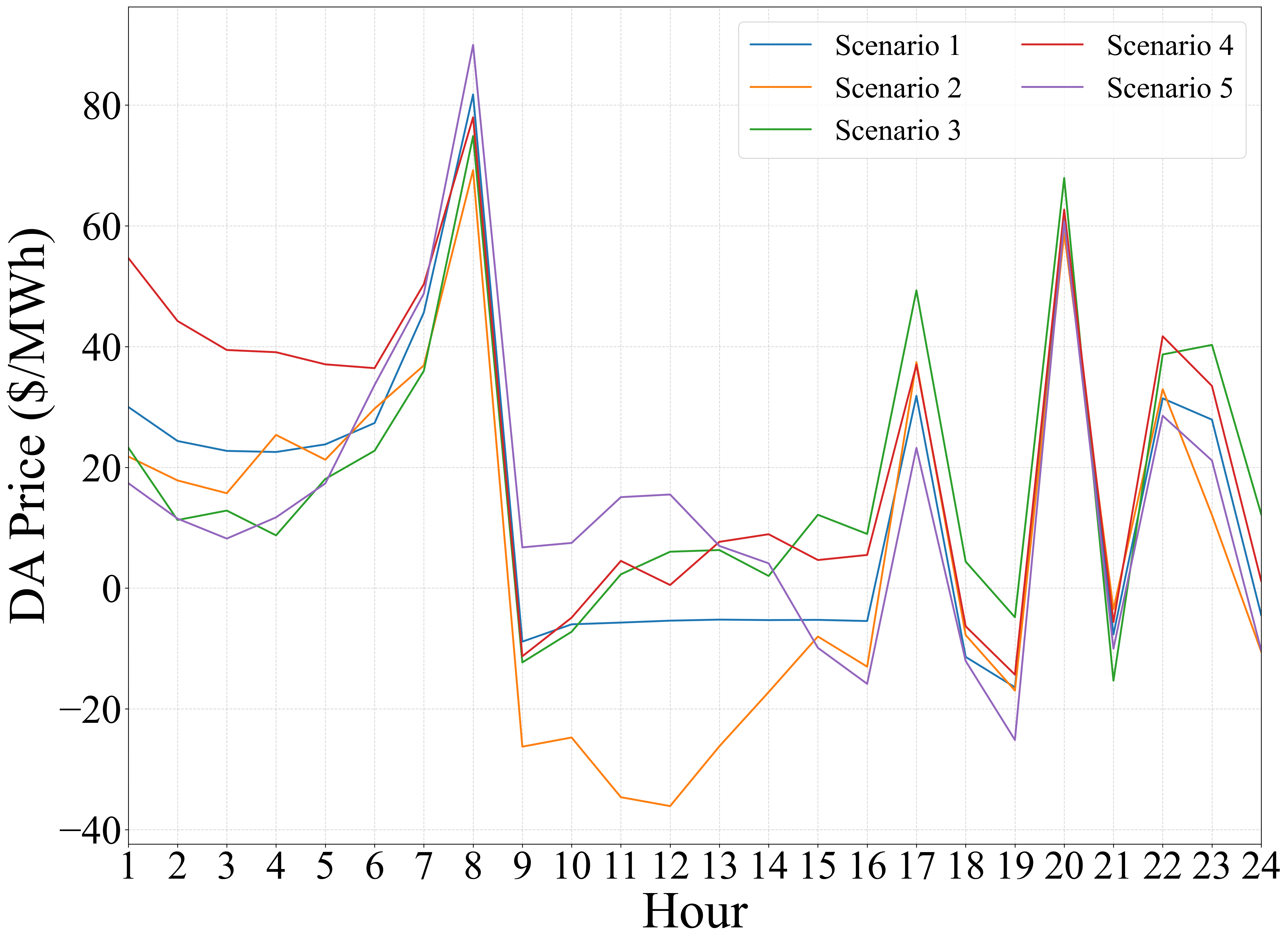}}
    \subfigure[]{\label{fig:syn_price_high}\includegraphics[width=0.48\linewidth]{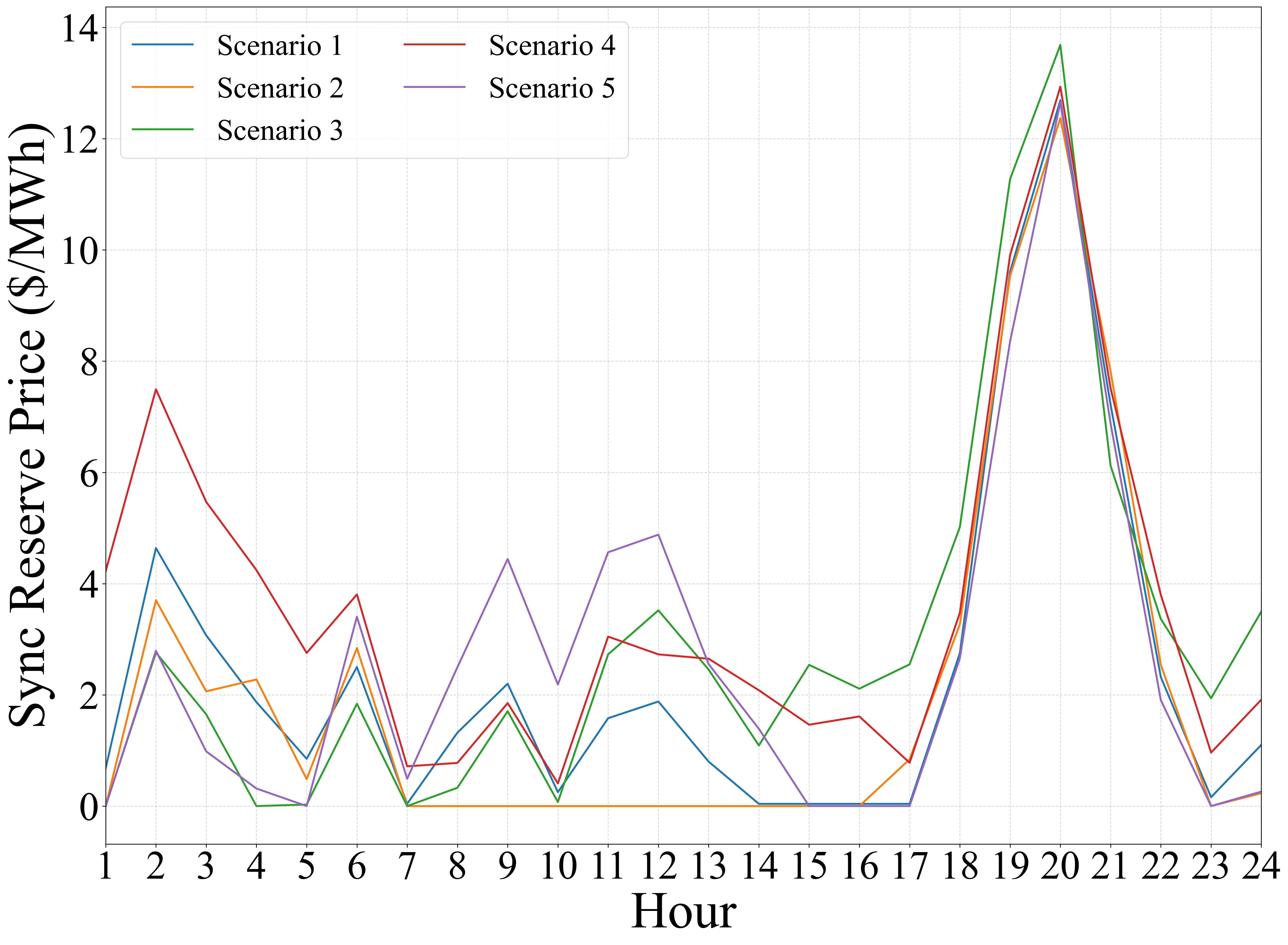}}
    \caption{(a) Day-ahead and (b) synchronized reserve price of case \textit{HIGH} for each scenario (both in \$/MWh).}
    \label{fig:price_high}
\end{figure}

{\KB First, the case referred to as \textit{HIGH} demonstrates a scenario in which the energy price exhibits relatively large fluctuations, ranging approximately \$-16.4 to \$81.8/MWh. The price scenarios for the case are shown in Fig.~\ref{fig:price_high}. The synchronized reserve trajectories are also shown for completeness.}

{\KBB Table~\ref{tab:operating_modes_high} presents the optimal operating mode for each hour. In case \textit{HIGH}, the VS-PSH plant is active for 7~h (hours 6, 18, 19, 20, 21, 22, and 24), with `Discharging' mode at hours 6 and 22 and `Continuous' mode at hours 18, 19, 20, 21, and 24.}

\begin{table}[t!]
\centering
\caption{{\KB Optimal operating mode of case \textit{HIGH} by hour}}
\label{tab:operating_modes_high}
\begin{tabular}{cccccc}
\toprule
\textbf{Hour} & \textbf{Mode} &
\textbf{Hour} & \textbf{Mode} &
\textbf{Hour} & \textbf{Mode} \\ \midrule

1  & Non-operating &
2  & Non-operating &
3  & Non-operating \\

4  & Non-operating &
5  & Non-operating &
6  & Non-operating \\

7  & Non-operating &
8  & Non-operating &
9  & Non-operating \\

10 & Non-operating &
11 & Non-operating &
12 & Non-operating \\

13 & Non-operating &
14 & Non-operating &
15 & Non-operating \\

16 & Non-operating &
17 & Non-operating &
18 & Continuous    \\

19 & Continuous    &
20 & Continuous    &
21 & Continuous    \\

22 & Discharging   &
23 & Non-operating &
24 & Continuous    \\ \bottomrule
\end{tabular}
\end{table}

\begin{figure*}[t!]
    \centering
    \includegraphics[width=\linewidth]{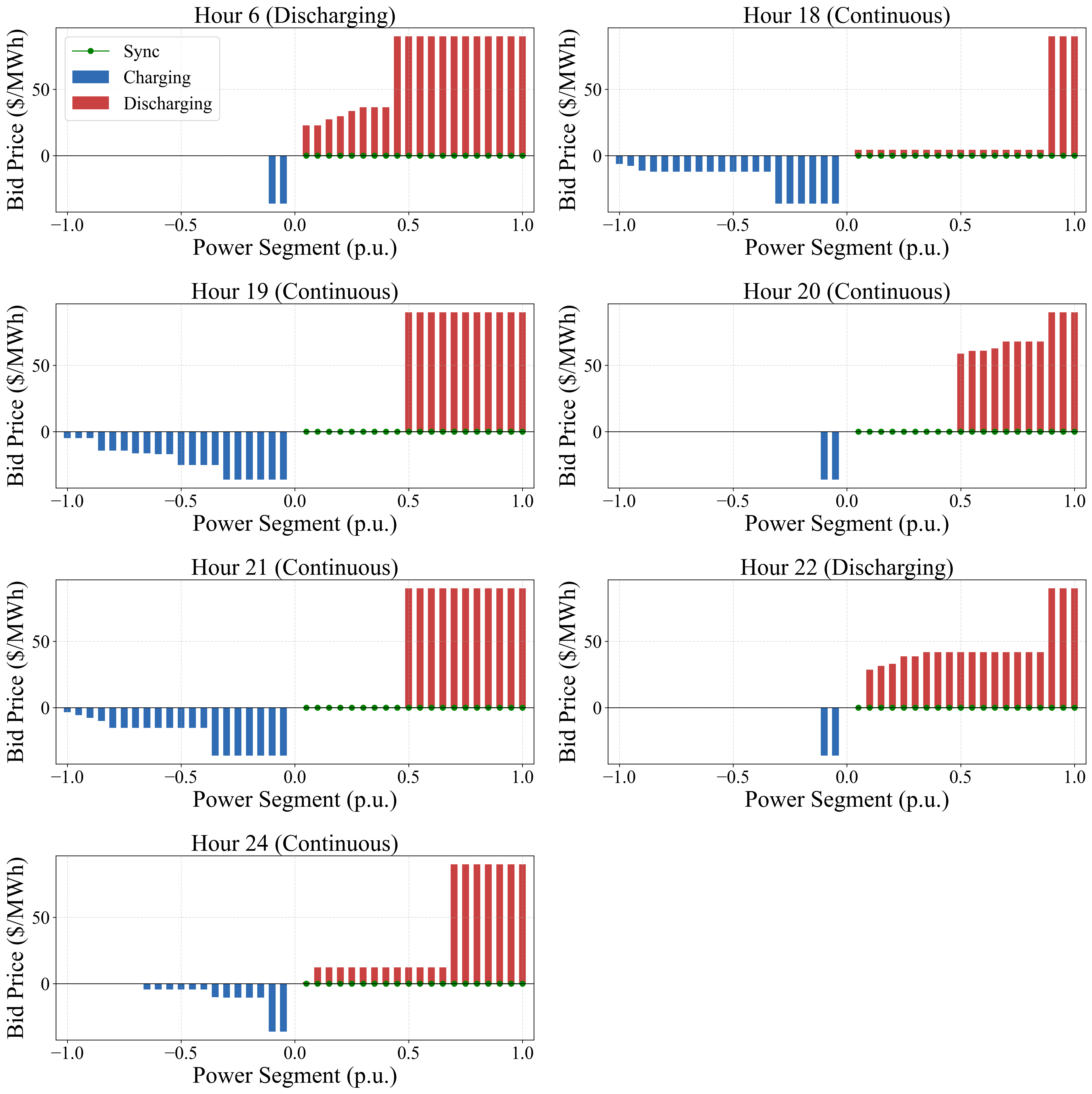}  
    \caption{Bidding result of case \textit{HIGH} for active operating hours (bid prices in \$/MWh, power segments in p.u.).}
    \label{fig:bidding_results_high} 
\end{figure*}

\begin{table}[t]
\centering
\caption{{\KB Expected hourly profit and cost of case \textit{HIGH}}}
\label{tab:hourly_summary_high}
\begin{tabular}{@{}>{}C{0.8cm} >{}R{2.5cm} >{}R{2.5cm} >{}R{2.5cm} >{}R{3cm}@{}}
\toprule
\textbf{Hour} & \textbf{Energy Revenue (\$)} & \textbf{Charging Cost (\$)} & \textbf{Sync Reserve Revenue (\$)} & \textbf{Total Hourly Profit (\$)} \\ \midrule
6              & 839.6                   & 0.0                     & 22.1                           & 861.7                        \\
18             & 0.7                     & -2000.7                 & 28.0                           & -1972.0                      \\
19             & 0.0                     & -2112.2                 & 89.4                           & -2022.8                      \\
20             & 7348.9                  & 0.0                     & 37.2                           & 7386.1                       \\
21             & 0.0                     & -1391.5                 & 70.1                           & -1321.4                      \\
22             & 998.3                   & 0.0                     & 22.5                           & 1020.9                       \\
24             & 4.2                     & -352.1                  & 10.9                           & -337.0                       \\ \midrule
Total          & 9191.8                  & -5856.5                 & 280.2                          & 3615.5                       \\
\bottomrule
\end{tabular}
\end{table}

\begin{figure}[t!]
    \centering
    \includegraphics[width=0.8\linewidth]{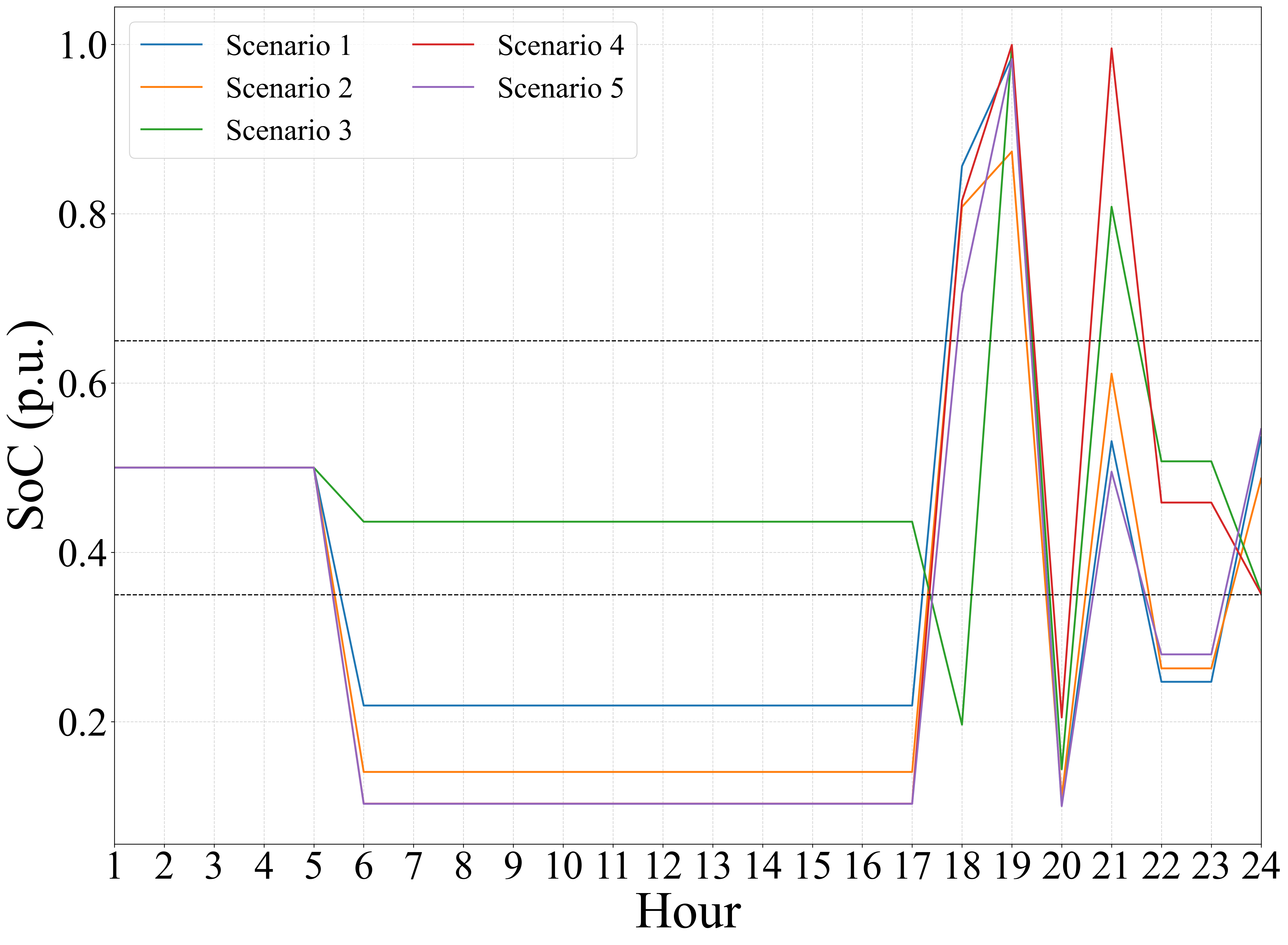}  
    \caption{SoC trajectory (p.u.) of case \textit{HIGH} for each scenario.}
    \label{fig:soc_trajectory_high} 
\end{figure}

{\KBB For the active hours identified in Table~\ref{tab:operating_modes_high}, the framework generates detailed bidding curves for the energy market, as illustrated in Fig.~\ref{fig:bidding_results_high}. Each subplot displays the multi-step charging (blue bars) and discharging (red bars) bids, and synchronized reserve bids are maintained at the minimum allowable price (\$0.01), preserving reserve-market accessibility while the model optimizes energy-reserve co-participation. In this updated case, major positive contributions occur at hours 20 and 22, while hours 18 and 19 are used for intertemporal SoC repositioning under uncertainty.}

{\KBB The financial outcome is summarized in Table~\ref{tab:hourly_summary_high}. After applying the requested scaling to energy, charging, and synchronized-reserve terms, the total expected daily profit is \$3615.5, with the largest positive contribution at hour 20 (\$7386.1). Negative-profit hours (notably hours 18, 19, 21, and 24) reflect intertemporal SoC repositioning decisions that improve overall horizon-level performance.} {\KBB In particular, the loss at hour~21 ($-$\$1,321.4) is driven primarily by a charging cost ($-$\$1,391.5): following the large discharge at hour~20, the optimizer deliberately recharges to restore the reservoir head---enabling physically feasible discharging at hour~22 under the head-dependent coupling constraint~\eqref{eq:integration_b}---and to prevent a violation of the SoC lower bound~\eqref{eq:soc_b} toward the end of the scheduling horizon.}

{\KB Fig.~\ref{fig:soc_trajectory_high} illustrates the SoC trajectory for each reduced scenario. The SoC evolves according to the cleared charging/discharging outcomes and remains within the prescribed operational limits throughout the 24-h horizon, confirming feasibility of the optimized schedule under uncertainty.}

\subsection{Case Study \#2: Case \textit{LOW}}

\begin{figure}[b!]
    \centering
    \subfigure[]{\label{fig:da_price_low}\includegraphics[width=0.48\linewidth]{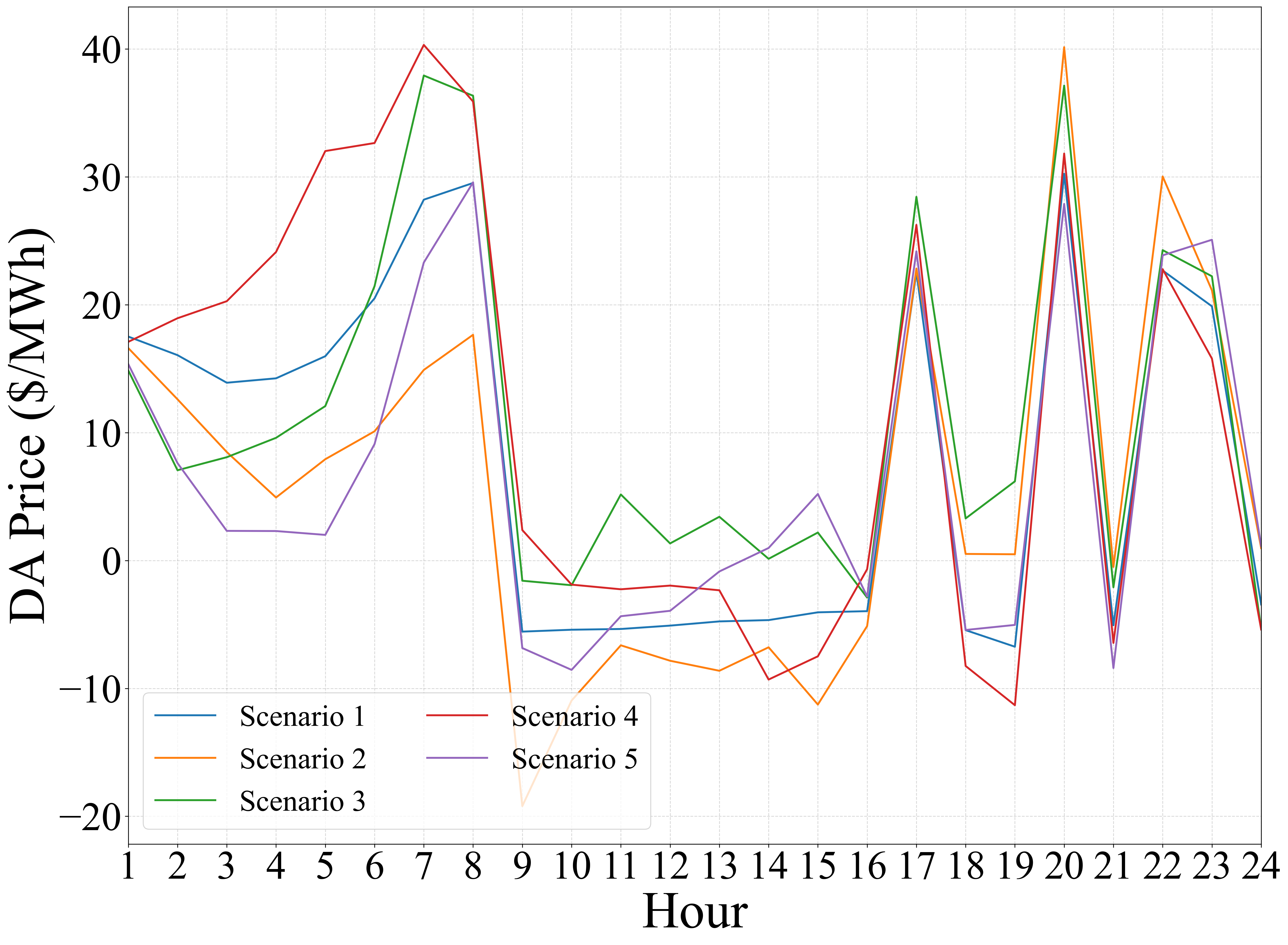}}
    \subfigure[]{\label{fig:syn_price_low}\includegraphics[width=0.48\linewidth]{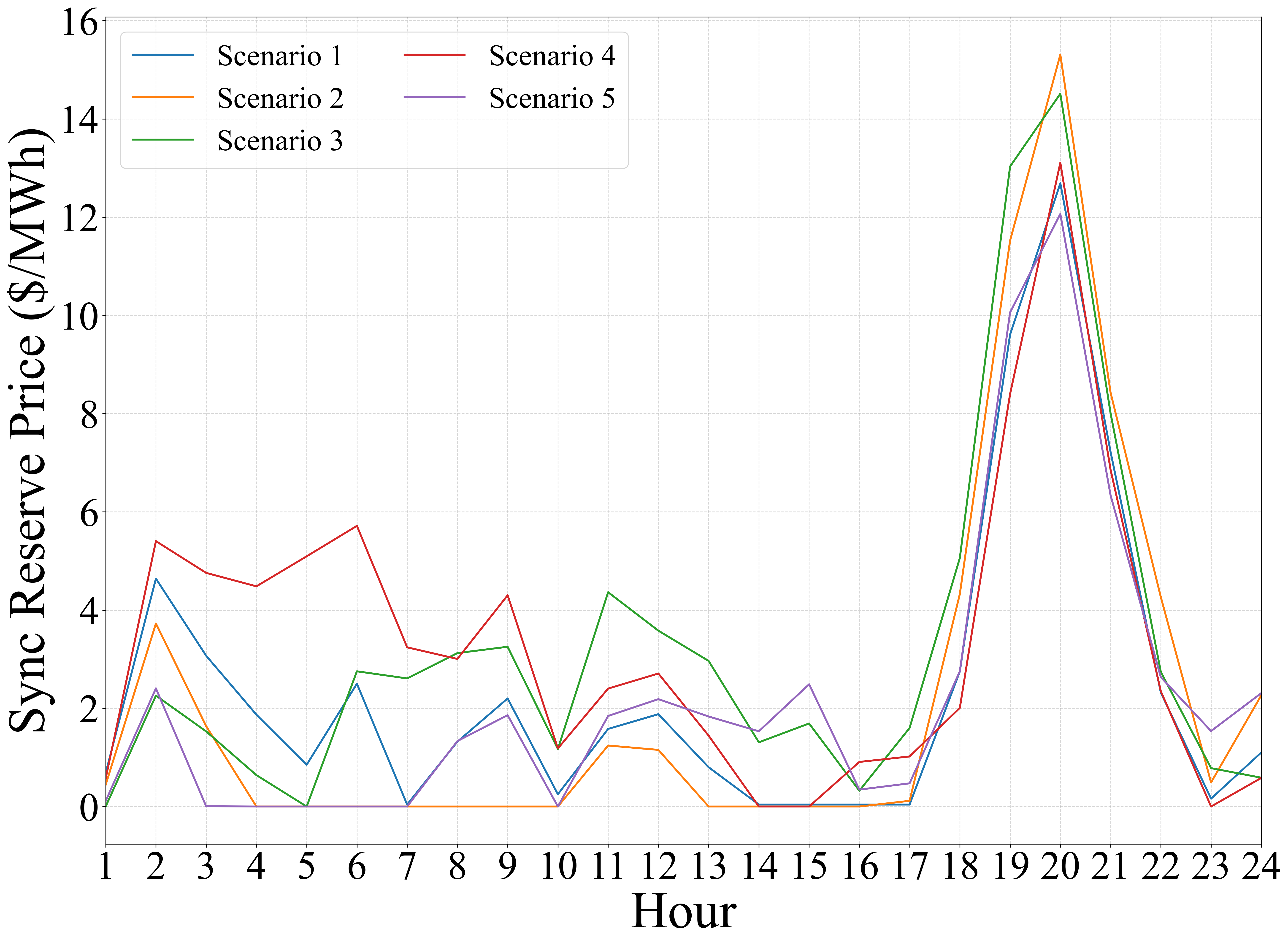}}
    \caption{(a) Day-ahead and (b) synchronized reserve price of case \textit{LOW} for each scenario (both in \$/MWh).}
    \label{fig:price_low}
\end{figure}

\begin{table}[b!]
\centering
\caption{{\KB Optimal operating mode of case \textit{LOW} by hour}}
\label{tab:operating_modes_low}
\begin{tabular}{cccccc}
\toprule
\textbf{Hour} & \textbf{Mode} &
\textbf{Hour} & \textbf{Mode} &
\textbf{Hour} & \textbf{Mode} \\ \midrule

1  & Non-operating &
2  & Non-operating &
3  & Non-operating \\

4  & Non-operating &
5  & Non-operating &
6  & Non-operating \\

7  & Non-operating &
8  & Non-operating &
9  & Non-operating \\

10 & Charging      &
11 & Non-operating &
12 & Non-operating \\

13 & Non-operating &
14 & Non-operating &
15 & Non-operating \\

16 & Non-operating &
17 & Continuous    &
18 & Non-operating \\

19 & Continuous    &
20 & Discharging   &
21 & Continuous    \\

22 & Continuous    &
23 & Non-operating &
24 & Continuous    \\ \bottomrule
\end{tabular}
\end{table}

\begin{figure*}[t!]
    \centering
    \includegraphics[width=\linewidth]{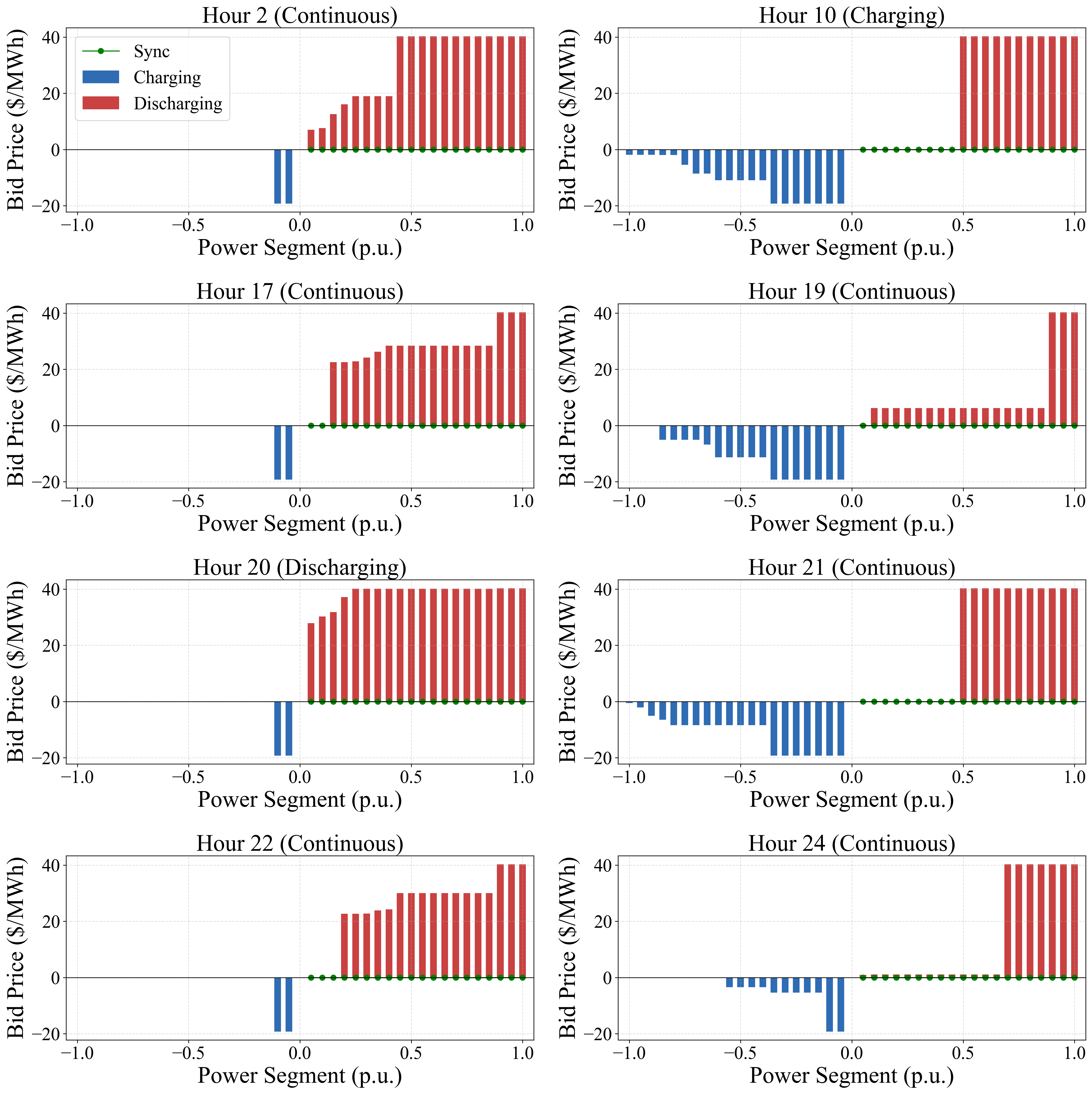}  
    \caption{Bidding result of case \textit{LOW} for active operating hours (bid prices in \$/MWh, power segments in p.u.).}
    \label{fig:bidding_results_low} 
\end{figure*}

\begin{figure}[t!]
    \centering
    \includegraphics[width=0.8\linewidth]{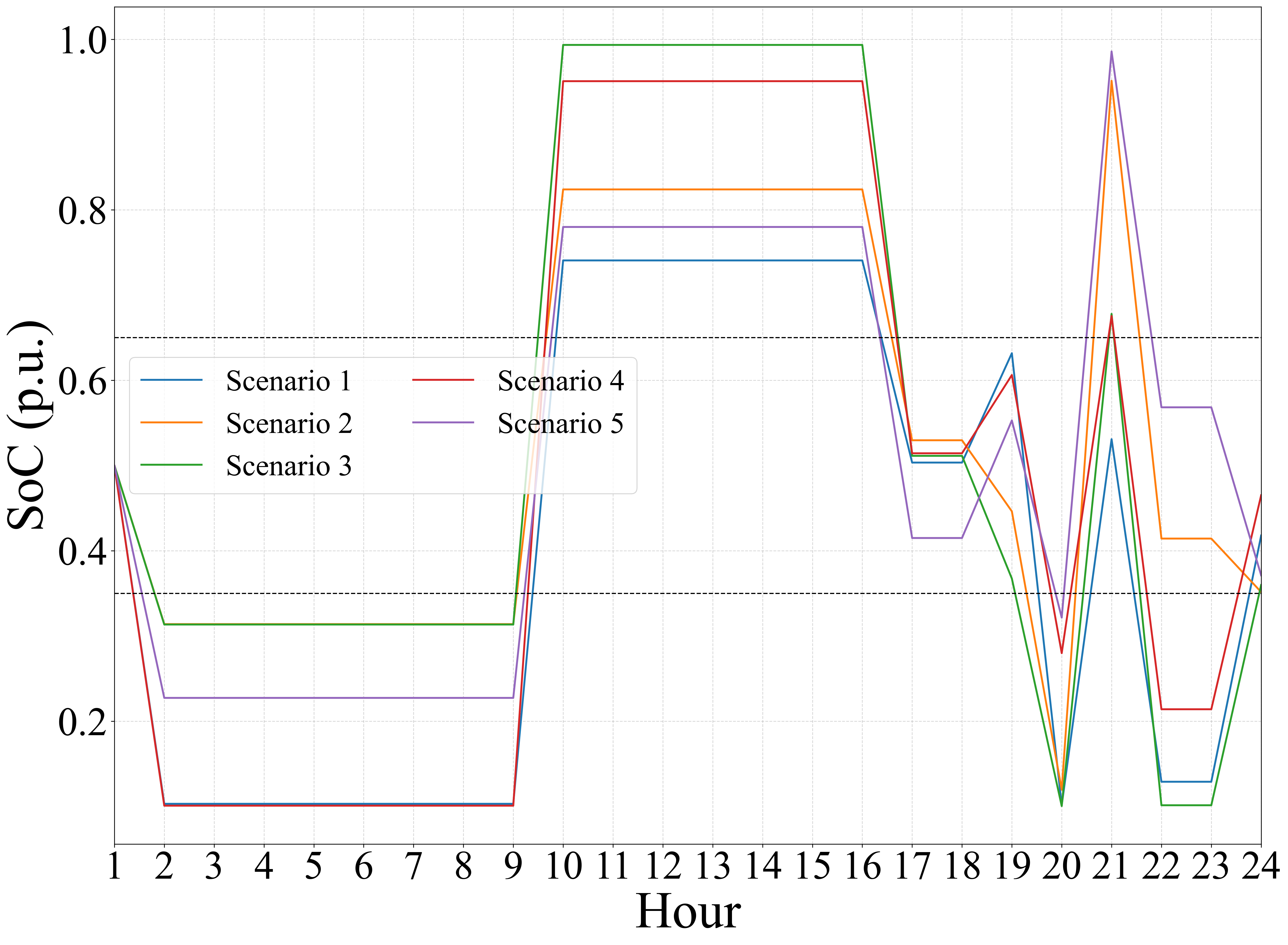}  
    \caption{SoC trajectory (p.u.) of case \textit{LOW} for each scenario.}
    \label{fig:soc_trajectory_low} 
\end{figure}

\begin{table}[t!]
\centering
\caption{{\KB Expected hourly profit and cost of case \textit{LOW}}}
\label{tab:hourly_summary_low}
\begin{tabular}{@{}>{}C{0.8cm} >{}R{2.5cm} >{}R{2.5cm} >{}R{2.5cm} >{}R{3cm}@{}}
\toprule
\textbf{Hour} & \textbf{Energy Revenue (\$)} & \textbf{Charging Cost (\$)} & \textbf{Sync Reserve Revenue (\$)} & \textbf{Total Hourly Profit (\$)} \\ \midrule
2  & 634.1   & 0.0      & 33.8   & 667.9     \\
10 & 0.0     & -808.9   & 0.0    & -808.9    \\
17 & 939.8   & 0.0      & 0.6    & 940.5     \\
19 & 1.5     & -858.9   & 88.9   & -768.6    \\
20 & 642.1   & 0.0      & 85.3   & 727.5     \\
21 & 0.0     & -898.4   & 70.4   & -828.0    \\
22 & 1176.1  & 0.0      & 22.7   & 1198.8    \\
24 & 0.3     & -270.7   & 10.7   & -259.8    \\ \midrule
Total & 3393.9 & -2836.9 & 312.5  & 869.5     \\
\bottomrule
\end{tabular}
\end{table}

{\KB To further test the adaptability of the framework, we conducted a second case study on case \textit{LOW}, which exhibits lower energy-price variation than case \textit{HIGH}. In the base scenario, the energy price in case \textit{LOW} ranges approximately from \$-6.7 to \$30.2/MWh, whereas case \textit{HIGH} reaches approximately \$-16.4 to \$81.8/MWh. The synchronized reserve profiles are generated from the same reserve dataset and reduction pipeline, so the key difference between the two cases is primarily driven by the energy-price dynamics.}

{\KBB This difference in market conditions changes the operation schedule. Table~\ref{tab:operating_modes_low} shows that case \textit{LOW} is active for 8~h (hours 2, 10, 17, 19, 20, 21, 22, and 24), with `Charging' mode at hour 10, `Discharging' mode at hour 20, and `Continuous' mode in the remaining active hours. The corresponding bidding patterns are shown in Fig.~\ref{fig:bidding_results_low}.}

{\KBB In terms of financial performance, Table~\ref{tab:hourly_summary_low} reports a total expected profit of \$869.5 for case \textit{LOW}, which is lower than the \$3615.5 obtained in case \textit{HIGH}. Energy-market revenue is lower in case \textit{LOW} (\$3393.9 vs. \$9191.8), while synchronized reserve revenue is higher (\$312.5 vs. \$280.2). This indicates that the optimizer relies more heavily on reserve-related value in the lower-volatility energy-price environment.}

{\KB The SoC trajectories in Fig.~\ref{fig:soc_trajectory_low} remain within the operational bounds for all reduced scenarios over the 24-h horizon, consistent with case \textit{HIGH}. This confirms that the proposed framework maintains feasibility and profitability across different price-volatility regimes.}

\subsection{Sensitivity Test Result}
\begin{figure*}[t!]
    \centering
    \includegraphics[width=\textwidth]{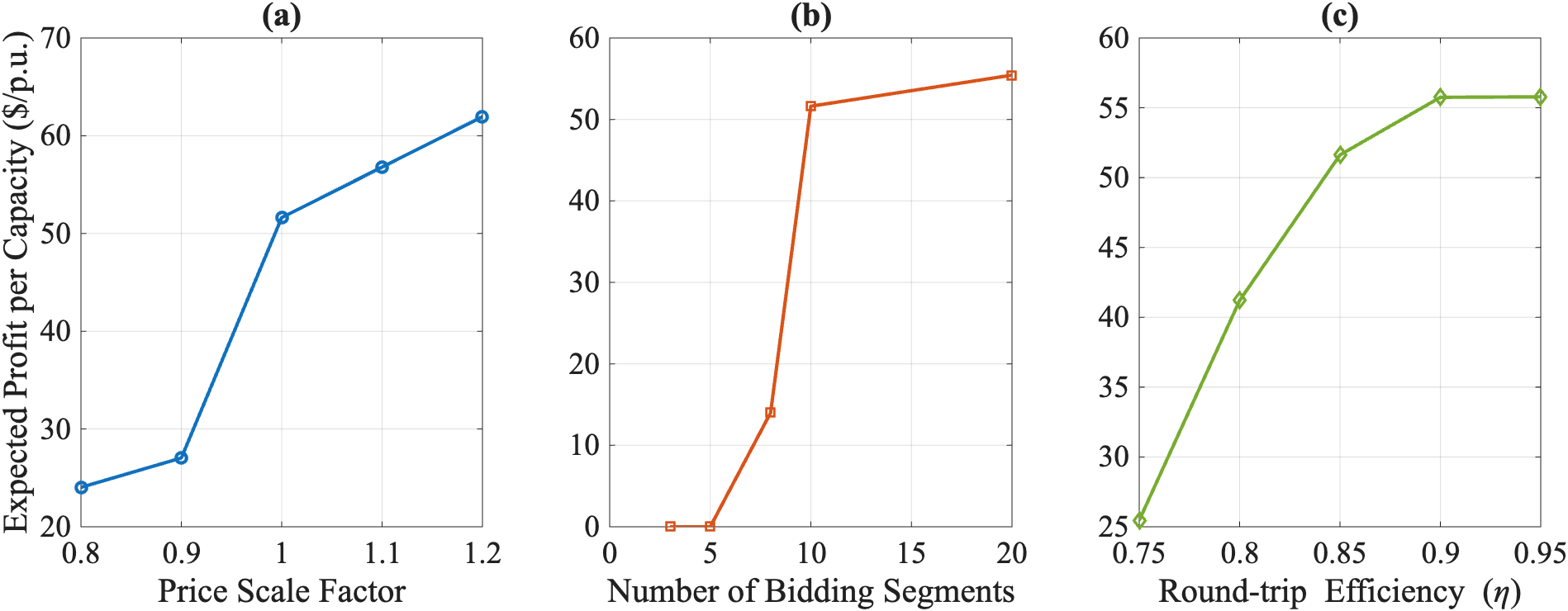} 
    \caption{Sensitivity analysis of the expected profit per unit capacity with respect to changes in (a) the market price scale factor, (b) the number of bidding segments, and (c) the round-trip efficiency ($\eta$).}
    \label{fig:sensitivity_analysis}
\end{figure*}

To evaluate the robustness of the proposed scheduling framework and understand how its performance is influenced by key parameters, a one-at-a-time (OAT) sensitivity analysis was conducted. In this analysis, individual parameters were varied across a predefined range while all other parameters were held at their baseline values. The analysis focuses on the change in the primary objective: the total expected profit, normalized by the plant's rated capacity and reported as \$/p.u. over the 24-hour horizon. The results for three critical parameters—market price scale, number of bidding segments, and round-trip efficiency—are visualized in Fig.~\ref{fig:sensitivity_analysis}.

{\KBB The tested parameter ranges were selected based on practical operation and market-screening considerations. The market-price scale factor range (0.8--1.2) represents moderate downside/upside day-ahead volatility around the baseline profile. The round-trip efficiency range (0.75--0.95) spans typical engineering values used for pumped-storage studies from conservative to high-performance operation. For bid granularity, we examined $n^{\mathrm{bid}} \in \{3,5,8,10,20\}$ to cover coarse, medium, and fine offer discretizations; $n^{\mathrm{bid}}=10$ is used as the baseline because it captures most profitability gain while limiting MILP size growth. In the stochastic setup, $N_s=5$ representative scenarios were retained after fast-forward reduction from $N_{\mathrm{gen}}=360$, balancing uncertainty representation and computational tractability. The synchronized-reserve floor bid of \$0.01 is used as a minimum non-zero offer to preserve reserve-market eligibility while avoiding artificial inflation of reserve-bid prices.}

\textbf{Impact of Price Scale Factor:} As shown in Fig.~\ref{fig:sensitivity_analysis}(a), the expected profit per capacity exhibits a strong positive correlation with the market price scale factor. When market prices are scaled down by 20\% (a factor of 0.8), the profit is approximately \$24/p.u.; when scaled up by 20\% (a factor of 1.2), the profit increases to over \$61/p.u. This result validates that the model effectively capitalizes on market volatility. Higher price levels and wider price spreads between charging and discharging periods directly translate to greater arbitrage opportunities and, consequently, higher profitability.

\textbf{Impact of Number of Bidding Segments:} The number of segments used to construct the bidding curves has a significant, non-linear effect on profitability, as illustrated in Fig.~\ref{fig:sensitivity_analysis}(b). With too few segments (e.g., 3 or 5), the model fails to generate any profit, indicating that the coarse bidding structure is insufficient to capture profitable arbitrage opportunities. As the number of segments increases to 8 and 10, the profit rises sharply to \$14/p.u. and \$51.6/p.u., respectively. A further increase to 20 segments yields only a marginal profit gain (to \$55.4/p.u.). This highlights a critical trade-off between model fidelity and computational complexity, suggesting that 10 segments offer a good balance for practical applications.

\textbf{Impact of Round-trip Efficiency ($\eta$):} Fig.~\ref{fig:sensitivity_analysis}(c) demonstrates that round-trip efficiency is a crucial driver of profitability. The expected profit per capacity grows substantially from approximately \$25/p.u. at an efficiency of 75\% to over \$55/p.u. at 95\%. Higher efficiency reduces energy losses during the charge-discharge cycle, making each arbitrage transaction more profitable and underscoring the model's logical response to the physical characteristics of the energy storage asset.

In summary, the sensitivity analysis confirms that the proposed framework is robust and responds logically to changes in market conditions, bidding structure, and physical characteristics. The results underscore the importance of market price volatility and plant efficiency in determining profitability, while also revealing a key trade-off between bidding granularity and model performance.

\subsection{Additional Comparative and Computational Analysis}
{\KB To address the reviewer request for quantitative benchmarking, we conducted additional ablation runs using the same 24-h horizon, scenario set ($N_s=5$), and solver settings. Table~\ref{tab:ablation_benchmark} compares (i) a single-segment benchmark, (ii) the proposed multi-segment head-dependent model, and (iii) a head-relaxed benchmark (proxy for a model without head-dependent capability envelopes, implemented by replacing head-dependent limits with static power bounds).}


\begin{table}[t!]
\centering
\caption{{\KB Quantitative benchmark across bidding granularity and head modeling assumptions}}
\label{tab:ablation_benchmark}
\begin{tabular}{@{}>{\centering\arraybackslash}m{3.5cm} >{\centering\arraybackslash}m{1cm} >{\centering\arraybackslash}m{2cm} >{\centering\arraybackslash}m{2.3cm} >{\centering\arraybackslash}m{2.9cm}}
\toprule
Configuration & $n^{bid}$ & Head-dependant & Expected Profit (\$/p.u.) & Head-Feasibility Violation Rate \\ \midrule
Single-segment benchmark & 1 & \checkmark & 0.00 & 0.0\% \\
Proposed model & 10 & \checkmark & 44.48 & 0.0\% \\
Head-relaxed benchmark & 10 & \xmark & 53.72 & 29.2\%\\ 
\bottomrule
\end{tabular}
\end{table}

\begin{table}[t!]
\centering
\caption{{\KB Computational complexity versus number of bid segments}}
\label{tab:complexity_segments}
\begin{tabular}{@{}>{}c>{}c>{}c>{}c>{}c@{}}
\toprule
$n^{\mathrm{bid}}$ & Variables & Constraints & Solve time (s) & Explored nodes \\ \midrule
3  & 3437  & 13335  & 0.014 & 0 \\
5  & 5021  & 22959  & 0.018 & 0 \\
8  & 7397  & 40095  & 0.250 & 1 \\
10 & 8981  & 53319  & 1.144 & 3442 \\
20 & 16901 & 141039 & 8.939 & 46006 \\ 
\bottomrule
\end{tabular}
\end{table}

{\KB The ablation results show that moving from a single-segment bid to the proposed multi-segment structure yields a large profitability gain while preserving physical feasibility. The head-relaxed benchmark gives a higher nominal profit, but nearly one-third of active scenario-hour points violate the original head-feasibility envelope, indicating that this additional profit is not physically realizable under plant constraints.} {\KBA Here, the head-feasibility violation rate is defined as the fraction of active scenario-hour pairs $(t,s)$ in which the dispatched power falls outside $\mathcal{F}^{ds}(h_{t,s})$ or $\mathcal{F}^{ch}(h_{t,s})$ when evaluated against the original head-dependent limits after removing those constraints from the benchmark.}

{\KB We also quantified computational complexity versus bid granularity. Table~\ref{tab:complexity_segments} reports the MILP scale and solve effort for representative segment counts.}
{\KB The results confirm the expected trade-off: finer bidding granularity improves objective value but increases MILP size and branch-and-bound effort nonlinearly. In our case, $n^{\mathrm{bid}}=10$ provides most of the economic benefit with substantially lower runtime than $n^{\mathrm{bid}}=20$, supporting its use as the baseline setting.}


{\KB \section{Discussion and Conclusions} \label{sec:CON}}

{\KB \subsection{Discussion}}
{\KB This study adopts a constant efficiency parameter for the SoC dynamics. This choice is motivated by the nature of the day-ahead bidding problem: the actual operating point is determined ex-post by market clearing, limiting the practical benefit of operating-point-dependent efficiency at the bidding stage. Moreover, the dominant effect of head variation on VS-PSH capability is already captured through the head-dependent feasible sets in the proposed framework. The sensitivity analysis in Section 4.3 confirms that mode selection decisions are robust to efficiency perturbations, with the qualitative scheduling pattern remaining consistent across the tested range. Dynamic efficiency formulations and the constant efficiency approach serve complementary purposes: the former suits dispatch-level optimization where operating points are known, while the latter prioritizes computational tractability for market-oriented bidding under stochastic uncertainty.}

{\KB \subsection{Conclusions}}
This study developed and validated a day-ahead, market-oriented scheduling framework for VS-PSH that explicitly captures key physical constraints while producing market-compatible multi-segment bids for both energy and synchronized reserves. By formulating the offering problem as a stochastic MILP and employing a compact scenario-generation procedure, the framework delivers feasible hourly mode schedules and segmented bidding curves that better reflect the marginal cost/benefit of charging, discharging, and reserve provision. Numerical results from cases \textit{HIGH} and \textit{LOW} indicate that the proposed approach performs profitable arbitrage under price uncertainty, respects head- and SoC-related operational limits across scenarios, and can opportunistically earn reserve revenue in addition to energy market gains. These findings suggest that VS-PSH owners can improve market outcomes and system flexibility by adopting multi-segment, co-optimized offering strategies that respect device physics and market compliance.

Future work will pursue {\KB three} complementary directions to increase realism and market relevance. First, we will strengthen the uncertainty treatment by developing larger, statistically calibrated scenario trees and/or robust-optimization formulations that capture extreme events and tail risk. Second, we will integrate market-interaction features by coupling the VS-PSH scheduling model with network-aware market clearing. 
{\KB Third, we will explore the integration of piecewise linear efficiency approximations within the MILP framework to capture dynamic head--flow--efficiency relationships for applications where detailed manufacturer characteristic curve data are available.}
These directions aim to bridge the gap between device-level physical realism and system-level market operations, improving decision support for VS-PSH owners and market operators alike.

%
\bibliographystyle{elsarticle-num}
\bibliography{references}
\itemsep2pt

\end{document}